\documentclass[12pt,a4paper]{iopart}

\usepackage{iopams}
\usepackage{setstack}
\usepackage{graphicx}
\usepackage{pict2e}
\usepackage{color}
\usepackage{hyperref}

\newcommand{\sgn}{\mathrm{sgn}}

\newcommand{\Li}{\mathrm{Li}}
\newcommand{\erf}{\mathrm{erf}}
\renewcommand{\Re}{\mathrm{Re}}
\renewcommand{\Im}{\mathrm{Im}}

\newcommand{\affiliation}[1]{\address{#1}}
\renewcommand{\pacs}[1]{\noindent\textbf{PACS numbers:} #1}
\newcommand{\keywords}[1]{\noindent\textbf{Keywords:} #1}
\newcommand{\tfrac}[2]{\mbox{\small$\frac{#1}{#2}$}}
\renewcommand{\text}[1]{\mathrm{#1}}

\begin{document}

\title[Perturbative solution for the spectral gap of WASEP]{Perturbative solution for the spectral gap of the weakly asymmetric exclusion process}
\author{Sylvain Prolhac}
\affiliation{Laboratoire de Physique Th\'eorique, IRSAMC, UPS, Universit\'e de Toulouse, CNRS, France}

\begin{abstract}
We consider the weakly asymmetric exclusion process with $N=L/2$ particles on a periodic lattice of $L$ sites, and hopping rates $1$ and $q=1-\mu/\sqrt{L}$ respectively in the forward and in the backward direction. Using Bethe ansatz, we obtain a systematic perturbative expansion of the spectral gap near $\mu=0$ by solving order by order a simple functional equation. A key point is that when $\mu\to0$, Bethe roots at a distance $1/\sqrt{L}$ from the edge of the Fermi sea should not be considered as a continuum, but converge instead at large $L$ to the complex zeroes of $1+\erf(x)$ after a rescaling by $\sqrt{L}$.\\

\\
\end{abstract}

\maketitle

\begin{section}{Introduction}
\label{section introduction}
The asymmetric simple exclusion process (ASEP) \cite{D1998.1,S2001.1,GM2006.1,S2007.1,CMZ2011.1} is a Markov process describing the dynamics of hard-core particles hopping on a lattice with a preferred direction. In one dimension, each particle can hop to the next site in the forward direction with rate $1$ and in the backward direction with rate $q<1$ if the destination site is empty. ASEP can be mapped to an interface growth model where elementary blocks deposit at rate $1$ on local minima $\vee$ and evaporate at rate $q$ from local maxima $\wedge$ of the interface.

ASEP belongs to the prominent non-equilibrium universality class known as KPZ \cite{KK2010.1,SS2010.4,C2011.1,T2014.1,QS2015.1,HHT2015.1,C2016.1,S2016.2,D2017.1}, from Kardar, Parisi and Zhang \cite{KPZ1986.1}, and which has been known to describes specific regimes of growing interfaces, driven lattice gases and directed polymers in random media. More recently, KPZ universality has been extended to several other settings featuring a strong interplay between noise and non-linearity, in particular one-dimensional systems with few conservation laws in the framework of non-linear fluctuating hydrodynamics \cite{S2014.1,S2016.1,PSSS2015.1} (including classical fluids \cite{vB2012.1}, anharmonic chains \cite{DDSMS2014.1,MS2015.1,SS2015.1} and quantum liquids described by the Gross-Pitaevskii equation \cite{KL2013.1,KHS2015.1}), as well as two-dimensional strongly localized systems \cite{SOP2007.1,SLDO2015.1} and free Fermions in a harmonic trap \cite{DLDMS2015.1,LDMRS2016.1}.

Under assumptions of local dynamics, sufficient non-linearity and uncorrelated noise, microscopic growth models are believed to flow quite generally at large scales to the universal KPZ fixed point. In the absence of non-linearity, the effective large scale dynamics is governed by the (repulsive) Edwards-Wilkinson fixed point. For the exclusion process, the KPZ and Edwards-Wilkinson fixed points are reached at large scales respectively for finite asymmetry ($q<1$) and symmetric hopping ($q=1$). Both fixed points are characterized by distinct universal fluctuations, Gaussian in the Edwards-Wilkinson case and non-Gaussian for KPZ. For specific initial conditions, fluctuations for infinitely long interfaces at the KPZ fixed point are described by Tracy-Widom distributions from random matrix theory, which have been observed in experiments on growing turbulent phases in liquid crystals \cite{TS2010.1,TSSS2011.1}.

The renormalization group flow connecting the Edwards-Wilkinson and the KPZ fixed point for any microscopic growth model with tunable asymmetry is also conjectured to be universal \cite{GJ2014.1,GP2015.1,HQ2015.1,HS2015.1}. This crossover is in particular realized \cite{BG1997.1,DM1997.1,SS2010.2,ACQ2011.1,GJ2014.1,GP2015.1,DT2016.1,CS2016.1} in the weakly asymmetric exclusion process (WASEP) with $1-q\sim\ell^{-1/2}$ after rescaling positions on the lattice by a factor $\ell$ and time by a factor $\ell^{2}$, $\ell\gg1$, see also \cite{H2014.1,HQ2015.1,HS2015.1} for another type of regularization of the universal crossover without lattice discretization. The crossover from Gaussian to KPZ fluctuations has been investigated in infinite systems in the droplet geometry \cite{SS2010.3,ACQ2011.1,D2010.1,CLDR2010.1}, as well as for flat \cite{CLD2011.1} and stationary \cite{IS2012.1} interfaces. Little is known so far about the crossover in finite systems, unlike at the KPZ fixed point where some progress happened recently \cite{P2016.1,BL2016.1,L2016.1,P2016.2} for the relaxation to the non-equilibrium steady state.

A distinguishing feature of ASEP among all microscopic models in KPZ universality is its stochastic integrability, a property shared with a small number of very specific models. Using Bethe ansatz, these models allow exact calculations of universal KPZ statistics, unlike more traditional perturbative or mean field approaches which usually fail due to the strongly coupled nature of the KPZ fixed point, see however \cite{CCDW2010.1,MAKLC2017.1} for recent advances using non-perturbative renormalization group methods. The generator $M$ of the Markov evolution $\rme^{tM}$ is equal to minus the Hamiltonian of a ferromagnetic XXZ spin chain with anisotropy $\Delta=(q^{1/2}+q^{-1/2})/2>1$ and twisted boundary conditions. The generator $M$ is non-Hermitian, except at $q=1$ where $-M$ reduces to the Hamiltonian of the ferromagnetic Heisenberg spin chain. Despite its integrability, ASEP is a faithful representative of KPZ universality at large scales. In particular, the hydrodynamic evolution does not feature infinitely many locally conserved quantities generated by a transfer matrix like in quantum integrable unitary evolutions: the density of particles in the exclusion process is the only independent conserved quantity since the transfer matrix is a Markov operator.

We consider in this paper the spectral gap of the generator $M$ in the weakly asymmetric regime with $q=1-\mu/\sqrt{L}$ for a system with $L$ sites and periodic boundary conditions, corresponding to an anisotropy $\Delta\simeq1+\mu^{2}/8L$ in the spin chain language. We study the limit of large system size $L$ and large number of particles $N$ with fixed density $\rho$, essentially $\rho=1/2$ for simplicity. The eigenvalue of $M$ with largest real part is equal to $0$ and corresponds to the stationary state of the system. The gap is equal to minus the real part of the second largest eigenvalue $E_{1}$. The imaginary part of $E_{1}$ scales as $L^{-3/2}$, with a coefficient proportional to the velocity $1-2\rho$ at which density fluctuations travel in the system. At half-filling $N=L/2$, the gap eigenvalue $E_{1}$ is real and behaves for large $L$ as $E_{1}\simeq e_{1}(\mu)L^{-2}$, corresponding to a relaxation time of order $L^{2}$ for the universal model identified by $\mu$ on the renormalization group flow between the Edwards-Wilkinson and KPZ fixed points.

By invariance of the problem under the replacement $q\to q^{-1}$, the function $e_{1}(\mu)$ is even. At $\mu=0$, the identification of $-M$ as the Hamiltonian of the ferromagnetic Heisenberg spin chain implies $E_{1}=-4\sin^{2}(\pi/L)$ for any $N$, which leads to $e_{1}(0)=-4\pi^{2}$. Close to $\mu=0$, diagonalization of $M$ for small system sizes provided the guess $e_{1}(\mu)\simeq-4\pi^{2}-\mu^{2}/2$ \cite{NdN1995.1}. Numerical investigations based on extrapolation of high precision Bethe ansatz numerics led to further conjectures for the coefficients of the expansion of $e_{1}(\mu)$ up to order $10$ in $\mu$ \cite{P2016.2}. At large $\mu$, the ASEP result $E_{1}\sim(1-q)L^{-3/2}$ leads for WASEP to $e_{1}(\mu)\sim\mu$. The coefficient, equal to $-6.509189\ldots$, was obtained by Gwa and Spohn \cite{GS1992.1,GS1992.2} for the totally asymmetric simple exclusion process (TASEP, $q=0$) using the special, almost decoupled nature of the Bethe equations in that case, see also \cite{GM2005.1,P2014.1} for a simpler derivation. Corrections of order $\mu^{-2}$ and $\mu^{-4}$ to $e_{1}(\mu)\sim\mu$ were obtained by Kim \cite{K1995.1} starting from the fully coupled Bethe equations in the case of general asymmetry $q$. The spectral gap of TASEP and ASEP with open boundary conditions \cite{dGE2005.1,dGE2006.1,dGE2008.1,dGFS2011.1} and with multiple species of particles \cite{AKSS2009.1,WK2010.1} have also been studied using Bethe ansatz.

The Bethe roots of WASEP can not conveniently be treated as a continuum. This is manifest in the approach followed by Kim in \cite{K1995.1} for the gap of ASEP, where the WASEP scaling limit re-sums contributions from all the orders of the large $L$ asymptotic expansion for ASEP. In this paper, we consider instead a rescaling of the Bethe roots at the edge of the Fermi sea in order to treat them in the large $L$ limit as an infinite number of discrete points that we call \textit{edge Bethe roots}. A crucial observation is that the Bethe roots in the bulk supply mostly divergent counter-terms making the contribution of the edge Bethe roots finite.

Several steps in our calculations are checked or even guessed using high precision numerics based on a numerical solution of Bethe equations with a few hundred digits combined with extrapolation methods. We applied extensively Richardson extrapolation \cite{R1927.1,BS1991.1}, a numerical technique to accelerate algebraic convergence of sequences with a single exponent, which is especially efficient in the study of integrable models \cite{P2016.2,P2017.1} since the exponent is usually known exactly in that case.

The paper is organized as follows. A key result, derived from a careful asymptotic evaluation of the Bethe equations in section \ref{section direct}, is that edge Bethe roots converge when $\mu\to0$ to the complex zeroes of $1+\erf(x)$. This is strongly reminiscent \footnote{We thank the anonymous referee for pointing that to us.} of results \cite{NR1972.1,V2015.1} about zeroes of truncated Taylor series of entire functions. The convergence of edge Bethe roots to the zeroes of $1+\erf(x)$ is then used in section \ref{section functional} to exhibit simple functional equations from which the small $\mu$ expansion of $e_{1}(\mu)$ can be computed systematically. The functional equations are obtained using the formulation of Bethe ansatz in terms of Baxter's TQ equation and of the quantum Wronskian, with the help of high precision numerics to overcome gaps in several steps of the derivation.
\end{section}

\begin{section}{Direct approach}
\label{section direct}
In this section, we study the Bethe roots of the weakly asymmetric exclusion process starting directly from the Bethe equations. Our main result, equation (\ref{Q[erf]}), is that edge Bethe roots (see definition below) converge to the zeroes of $1+\erf$ when $\mu\to0$.

\begin{subsection}{Bethe ansatz for ASEP}
Any eigenvector of the generator $M$ of ASEP can be constructed by the coordinate Bethe ansatz method as a linear combination of $N!$ permutations $\sigma$ of plane waves $z_{\sigma(1)}^{x_{1}}\ldots z_{\sigma(N)}^{x_{N}}$ with particles at positions $x_{j}$, $j=1,\ldots,N$ and coefficients depending on the complex numbers $z_{j}$, $j=1,\ldots,N$. Each eigenstate corresponds to a specific set of $N$ Bethe roots $z_{j}$ solution of coupled polynomial equations called the Bethe equations.

It is convenient to work with the Bethe roots $y_{j}=(1-z_{j})/(1-qz_{j})$ instead of the $z_{j}$'s. The Bethe equations are then
\begin{equation}
\label{BE[y]}
\Big(\frac{1-y_{j}}{1-qy_{j}}\Big)^{L}=(-1)^{N-1}\prod_{k=1}^{N}\frac{y_{j}-qy_{k}}{y_{k}-qy_{j}}\;,
\end{equation}
and the corresponding eigenvalue of $M$ is
\begin{equation}
\label{E[y]}
E=(1-q)\sum_{j=1}^{N}\Big(\frac{1}{1-y_{j}}-\frac{1}{1-qy_{j}}\Big)\;.
\end{equation}

Dividing both sides of (\ref{BE[y]}) by $y_{j}^{N}$, taking the logarithm and dividing the result by $L$ leads to the logarithmic form of the Bethe equations. The discontinuity of $2\rmi\pi$ across the branch cut of the logarithm implies that there must exist numbers $n_{j}$, $j=1,\ldots,N$ (integers for odd $N$ and half-integers for even $N$) such that
\begin{equation}
\label{LBE[y]}
f(y_{j};y_{1},\ldots,y_{N})=\frac{2\rmi\pi}{L}\,n_{j}\;,
\end{equation}
with $f$ the function defined by
\begin{equation}
\label{f[yj]}
f(y;y_{1},\ldots,y_{N})=\log\Big(\frac{1}{y^{\rho}}\,\frac{1-y}{1-qy}\Big)-\frac{1}{L}\sum_{j=1}^{N}\log_{j}\Big(\frac{1}{y}\,\frac{y-qy_{j}}{y_{j}-qy}\Big)\;.
\end{equation}
The function $\log_{j}$ is the logarithm with curved, $j$-dependent branch cut given by $\log_{j}(z)=\rmi\theta_{j}(|z|)+\log(\rme^{-\rmi\theta_{j}(|z|)}z)$, where $\log$ is the usual complex logarithm with branch cut $\mathbb{R}^{-}$: $\log(r\,\rme^{\rmi\varphi})=\log(r)+\rmi\varphi$ for $r>0$ and $-\pi<\varphi<\pi$. In order to make notations lighter, only the first argument of the function $f$ is written explicitly in the following.

In the special case $q=0$ of totally asymmetric dynamics, the logarithmic Bethe equations reduce to a kind of mean field problem, after introducing the variable $b=L^{-1}\sum_{j=1}^{N}\log y_{j}$ \cite{GS1992.1,DL1998.1,DA1999.1,PP2007.1}. This rewriting, combined with nice additional simplifications for the norm of eigenstates and overlaps \cite{B2009.1,MSS2012.1,MSS2012.2}, allowed precise asymptotic analysis \cite{P2015.2} and lead to explicit formulas for current fluctuations in finite volume \cite{P2016.1}, see also \cite{BL2016.1,BL2016.2,L2016.1} for an alternative approach based on the propagator.

For general $q$, the steady state corresponds to the singular solution $y_{j}=0$, which can be regularized by adding a twist counting the current of particles. With an appropriate choice for the angles $\theta_{j}$, the numbers $n_{j}$ then form the Fermi sea $n_{j}=j-\frac{N+1}{2}$. The spectral gap, which is doubly degenerate, corresponds in principle to the simplest particle-hole excitation over the Fermi sea, shifting either $n_{1}$ of $-1$ or $n_{N}$ of $+1$. Requiring convenient branch cuts for $f$ slightly complicates the matter. We choose $\theta_{j}(r)=-\arg y_{j}$, $j=1,\ldots,N-1$, independent of $r$, and $\theta_{N}(r)=-\pi+\arg\frac{r/y_{N}+q}{r/y_{N}+q^{-1}}$. The $N-1$ first angles $\theta_{j}$ avoid a discontinuity of $f(y)$ near $y=y_{j}$ and split the branch cut of $\log_{j}$ into two straight lines, one between $0$ and $qy_{j}$ and the other between $q^{-1}y_{j}$ and $+\infty y_{j}$, see figure \ref{fig branch cuts f}. The choice of $\theta_{N}$ imposes that the branch cut of $\log_{N}$ is also split into two parts, the one issuing from the branch point $0$ coinciding with the negative real axis, which is also the branch cut of $y^{\rho}=\rme^{\rho\log(y)}$, and the other one linking $qy_{N}$ and $q^{-1}y_{N}$.

With the previous definition for the $\theta_{j}$, we observe numerically on small systems (up to a few hundred sites) that the gap corresponds to $n_{j}=j-\frac{N-1}{2}$, $j=1,\ldots,N-1$, and $n_{N}$ a discontinuous function of $\mu$ depending on the disposition of $y_{N}$ compared with the branch cuts arising from the $q^{-1}y_{j}$, $j=1,\ldots,N-1$, see figure \ref{fig branch cuts f}.

\begin{figure}
  \begin{center}
    \includegraphics[width=75mm]{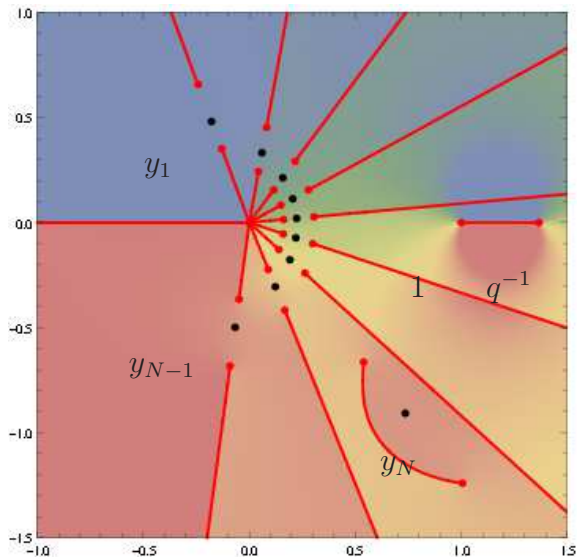}
    \begin{picture}(0,0)(4.5,5.5)
      \put(-19,40){$1$}
      \put(-9,40){$q^{-1}$}
      \put(-54,56.5){$y_{1}$}
      \put(-56,30){$y_{N-1}$}
      \put(-23,17){$y_{N}$}
    \end{picture}
  \end{center}
  \caption{Branch cuts of the function $f$ defined in (\ref{f[yj]}) for $\mu=1.2$, $L=20$, $N=10$. The branch cuts are represented by red, solid lines. The black dots are the Bethe roots $y_{j}$, while the red dots, from which the branch cuts originate, are the branch points $0$, $1$, $q^{-1}$, $qy_{j}$ and $q^{-1}y_{j}$, $j=1,\ldots,N$ of $f$. The background color represents the imaginary part of $f$, increasing from blue (upper left) to red (lower left).}
  \label{fig branch cuts f}
\end{figure}

Since $y_{N}$ is far from the other $y_{j}$'s, especially for small asymmetry $\mu$, our choice of the $\theta_{j}$'s clears a path along the $y_{j}$'s, $j=1,\ldots,N-1$ without branch cuts, as shown in figure \ref{fig branch cuts f}. For ASEP with finite asymmetry $1-q$, the path drawn by the Bethe roots $y_{j}$ stays at a finite distance from the branch points $qy_{j}$, $q^{-1}y_{j}$ of $f$. This allowed the calculation of the gap in \cite{K1995.1} by inserting $y_{j}=f^{-1}(2\rmi\pi n_{j}/L)$ in (\ref{f[yj]}) and replacing the sums by integrals plus boundary corrections using the Euler-Maclaurin formula. The problem is more complicated for WASEP due to the proximity between the path of the $y_{j}$'s and the surrounding branch points, leading to a pinching singularity in the middle of the integration range, corresponding in (\ref{f[yj]}) to $j$ such that $y_{j}\simeq y$.

In the rest of this section, we will be interested in the large $L$ expansion of the Bethe roots $y_{j}$, $y_{N-j}$ for fixed $j$, which requires the knowledge of $f$ at large $L$. For $y$ in the bulk of the path between $y_{1}$ and $y_{N-1}$, only the leading order will be needed for the following. For $y$ at a distance $\sim1/\sqrt{L}$ of the edges of the path, the first order correction will also be required.
\end{subsection}

\begin{subsection}{Asymptotics of \texorpdfstring{$f$}{f} in the bulk}
At leading order in $L$, the sum over $j$ in the definition (\ref{f[yj]}) of $f$ can be replaced by a Riemann integral, after writing $y_{j}=f^{-1}(\frac{2\rmi\pi}{L}(j-\frac{N-1}{2}))$ (the individual contribution of $y_{N}$ is negligible at this order). The leading order $f_{0}$ of $f$ thus verifies
\begin{eqnarray}
&&\hspace{-10mm} f_{0}(y)=\log\Big(\frac{1}{y^{\rho}}\,\frac{1-y}{1-qy}\Big)-\int_{-\rho/2}^{\rho/2}\rmd u\,\Big[-\rmi\arg(f_{0}^{-1}(2\rmi\pi u))\\
&&\hspace{55mm} +\log\Big(\rme^{\rmi\arg(f_{0}^{-1}(2\rmi\pi u))}\frac{1}{y}\,\frac{y-qf_{0}^{-1}(2\rmi\pi u)}{f_{0}^{-1}(2\rmi\pi u)-qy}\Big)\Big]\;.\nonumber
\end{eqnarray}
After the change of variables $z=f_{0}^{-1}(2\rmi\pi u)$, we obtain the integral equation
\begin{equation}
\label{f0[f0']}
\fl\hspace{10mm} f_{0}(y)=\log\Big(\frac{1}{y^{\rho}}\,\frac{1-y}{1-qy}\Big)-\oint_{\overline{\mathcal{C}}}\frac{\rmd z}{2\rmi\pi}\,f_{0}'(z)\Big[-\rmi\arg z+\log\Big(\rme^{\rmi\arg z}\frac{1}{y}\,\frac{y-qz}{z-qy}\Big)\Big]\;.
\end{equation}
Numerics indicate that the contour $\overline{\mathcal{C}}$ is closed and oriented clockwise, with $0$ lying inside the contour and $1$ outside. The solution to (\ref{f0[f0']}) is in fact independent of the asymmetry $q$ \cite{K1995.1}:
\begin{equation}
\label{f0}
f_{0}(y)=\rho\log\rho+(1-\rho)\log(1-\rho)+\log\Big(\frac{1-y}{y^{\rho}}\Big)\;.
\end{equation}
The curve $\{f_{0}^{-1}(2\rmi\pi u),-\rho/2<u<\rho/2\}$ to which the Bethe roots $y_{j}$ converge at large $L$ is plotted in figure \ref{fig curve y}. The point $-\rho/(1-\rho)=f_{0}^{-1}(\pm\rmi\pi\rho)$ corresponds to a square root singularity of $f_{0}$.

\begin{figure}
  \begin{center}
    \includegraphics[width=75mm]{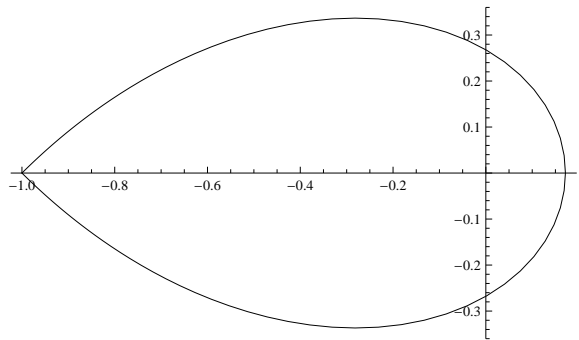}
  \end{center}
  \caption{Representation in the complex plane of the closed curve $f_{0}^{-1}(2\rmi\pi u)$, $-\frac{\rho}{2}<u<\frac{\rho}{2}$ on which the Bethe roots $y_{j}$ accumulate at large system size $L$ for a system with density $\rho=1/2$.}
  \label{fig curve y}
\end{figure}

The first correction of $f(y)$ to the leading term $f_{0}(y)$ is of order $1/L$. Inserting $y_{j}=f^{-1}(2\rmi\pi n_{j}/L)$ with $f(y)\simeq f_{0}(y)+f_{1}(y)/L$ into (\ref{f[yj]}) and expanding naively the sum up to first order in $L$ using the Euler-Maclaurin formula however only leads to the useless equality $f_{1}(y)=f_{1}(y)$. This is essentially what prevents us from performing the asymptotics of the gap using a direct approach from the Bethe equations, as one would need to know $f_{1}$ for that. The problem is circumvented in section \ref{section functional} by using a functional formulation. In the rest of this section, we study instead the first correction in $L$ of the Bethe roots $y_{j}$, $y_{N-j}$ for fixed $j$. The limit $\mu\to0$ of this correction will turn out to be a crucial ingredient in the functional approach of section \ref{section functional}.
\end{subsection}

\begin{subsection}{Asymptotics of \texorpdfstring{$f$}{f} at the edge}
\label{subsection f edge}
For fixed $j$, $y_{j}$ and $y_{N-j}$ converge to $f_{0}^{-1}(\pm\rmi\pi\rho)=-\rho/(1-\rho)$ at large $L$, with a correction of order $L^{-1/2}$ due to the square root singularity of $f_{0}$ at that point. We write
\begin{equation}
\label{y[w]}
\fl\hspace{10mm}
y_{j}=-\frac{\rho}{1-\rho}+\frac{w_{j}(\mu)}{\sqrt{L}}+\mathcal{O}\Big(\frac{1}{L}\Big)
\quad\text{and}\quad
y_{N-j}=-\frac{\rho}{1-\rho}+\frac{w_{-j}(\mu)}{\sqrt{L}}+\mathcal{O}\Big(\frac{1}{L}\Big)\;.
\end{equation}
The special Bethe root $y_{N}$ corresponds to $w_{0}(\mu)$. We call $w_{j}(\mu)$, $j\in\mathbb{Z}$ the edge Bethe roots. In the following, we frequently write $w_{j}$ instead of $w_{j}(\mu)$ to lighten notations when working with generic values of $\mu$.

We would like to have a system of equations characterizing the $w_{j}$'s, similar to the Bethe equations for the $y_{j}$'s. This requires the large $L$ expansion of $f(y)$ for $y=-\rho/(1-\rho)+w/\sqrt{L}$ up to order $L^{-1}$. Starting again with (\ref{f[yj]}), we split the sum over $j$ as
\begin{eqnarray}
\label{f[yj] splitted}
&&\fl\hspace{8mm} f(y)=\log\Big(\frac{1}{y^{\rho}}\,\frac{1-y}{1-qy}\Big)
-\frac{1}{L}\sum_{j=1}^{M}\log_{j}\Big(\frac{1}{y}\,\frac{y-qy_{j}}{y_{j}-qy}\Big)
-\frac{1}{L}\sum_{j=M+1}^{N-M-1}\log_{j}\Big(\frac{1}{y}\,\frac{y-qy_{j}}{y_{j}-qy}\Big)\nonumber\\
&&\hspace{25mm} -\frac{1}{L}\sum_{j=N-M}^{N-1}\log_{j}\Big(\frac{1}{y}\,\frac{y-qy_{j}}{y_{j}-qy}\Big)
-\frac{1}{L}\log_{N}\Big(\frac{1}{y}\,\frac{y-qy_{N}}{y_{N}-qy}\Big)\;,
\end{eqnarray}
and then take the large $L$ limit of each sum in (\ref{f[yj] splitted}) for $M\ll L$. The first and third sums can be treated easily by assuming (\ref{y[w]}). Setting $\rho=1/2$ from now on for simplicity, it leads to
\begin{eqnarray}
\label{f[yj] splitted 2}
&&\fl\hspace{13mm} f(y)\simeq-\frac{\rmi\pi}{2}\,\sgn(\Im w)+\frac{w+\mu}{2\sqrt{L}}+\frac{w^{2}+(w-\mu)^{2}}{8L}\nonumber\\
&& -\frac{1}{L}\sum_{j=1}^{M}\Big(-\rmi\pi+\log\frac{w_{j}-w+\mu}{w-w_{j}+\mu}\Big)-\frac{1}{L}\sum_{j=-M}^{-1}\Big(\rmi\pi+\log\frac{w_{j}-w+\mu}{w-w_{j}+\mu}\Big)\\
&& -\frac{1}{L}\Big(-\rmi\pi\,\sgn(\Im w)+\log\frac{w_{0}-w+\mu}{w-w_{0}+\mu}\Big)-\frac{1}{L}\sum_{j=M+1}^{N-M-1}\log_{j}\Big(\frac{1}{y}\,\frac{y-qy_{j}}{y_{j}-qy}\Big)\;.\nonumber
\end{eqnarray}
The last sum of (\ref{f[yj] splitted 2}), more complicated, is treated in \ref{appendix asymptotics second sum f}. There, we argue that when $M\gg1$, the Bethe root $y_{j}$ can be replaced at leading order by $f_{0}^{-1}(2\rmi\pi n_{j}/L)$. Using the Euler-Maclaurin formula with a careful treatment of the singularities at the edges, we obtain for $1\ll M\ll L$
\begin{eqnarray}
\label{asymptotics sum[logj,yj]}
&& -\frac{1}{L}\sum_{j=M+1}^{N-M-1}\log_{j}\Big(\frac{1}{y}\,\frac{y-qy_{j}}{y_{j}-qy}\Big)\simeq
-\frac{w+\mu}{2\sqrt{L}}
-\frac{w^{2}+\mu^{2}-w\mu/2}{4L}\\
&&\hspace{57mm} +\frac{\mu}{\sqrt{2\pi}L}\Big(-\zeta(1/2)+\sum_{j=1}^{M}\frac{1}{\sqrt{j}}\Big)\;.\nonumber
\end{eqnarray}
Gathering everything, we find eventually at large $L$
\begin{equation}
f(y)\simeq-\sgn(\Im w)\frac{\rmi\pi}{2}+\frac{g(w;\ldots,w_{-1},w_{0},w_{1},\ldots)}{L}\;,
\end{equation}
with
\begin{eqnarray}
\label{g(w)}
&&\fl g(w;\ldots,w_{-1},w_{0},w_{1},\ldots)=\sgn(\Im w)\rmi\pi-\frac{\mu(w+\mu)}{8}-\frac{\mu\,\zeta(1/2)}{\sqrt{2\pi}}-\log\frac{w_{0}-w+\mu}{w-w_{0}+\mu}\nonumber\\
&&\hspace{23mm} +\sum_{j=1}^{\infty}\Big(\frac{\mu}{\sqrt{2\pi j}}-\log\frac{w_{j}-w+\mu}{w-w_{j}+\mu}-\log\frac{w_{-j}-w+\mu}{w-w_{-j}+\mu}\Big)\;.
\end{eqnarray}
Numerics indicate that $w_{j}\simeq4\sqrt{\rmi\pi j}$ when $|j|\to\infty$, which ensures the convergence of the sum over $j$ in (\ref{g(w)}). In order to make notations lighter, only the first argument of the function $g$ is written explicitly in the following.

The edge Bethe roots $w_{j}$, $j\in\mathbb{Z}^{*}$ then verify the (logarithmic) edge Bethe equations
\begin{equation}
\label{BE[w]}
g(w_{j})=2\rmi\pi(j+\tfrac{1}{2})\;.
\end{equation}
The case of $w_{0}$ is more complicated because $f(y_{N})$ depends on $\mu$ through the position of $y_{N}$ with respect to the branch cuts of $f$. The issue disappears for small $\mu$, where we observe numerically that $w_{0}$ goes to infinity, which leaves coupled equations for the $w_{j}(0)$ with $j\in\mathbb{Z}^{*}$ only.
\end{subsection}

\begin{subsection}{Limit \texorpdfstring{$\mu\to0$}{mu->0}}
\label{subsection mu0}
In the limit $\mu\to0$, we observe numerically that $w_{0}(\mu)$ diverges as $w_{0}(\mu)\simeq-8\rmi\pi/\mu$, while all the other $w_{j}(\mu)$ converge to finite values $w_{j}(0)$, with the symmetry by complex conjugation $w_{-j}(0)=\overline{w_{j}(0)}$ for $j\in\mathbb{Z}^{*}$. From (\ref{g(w)}) and (\ref{BE[w]}), the edge Bethe equations become for all $j\in\mathbb{Z}^{*}$
\begin{equation}
\label{BE[w0]}
\frac{w_{j}(0)}{8}+\frac{\zeta(1/2)}{\sqrt{2\pi}}=\frac{1}{2\sqrt{\rmi\pi j}}+\sum_{k\in\mathbb{Z}^{*} \atop k\neq j}\Big(\frac{2}{w_{j}(0)-w_{k}(0)}+\frac{1}{2\sqrt{\rmi\pi k}}\Big)\;.
\end{equation}
We claim from (\ref{BE[w0]}) that the $w_{j}(0)$, $j\in\mathbb{Z}^{*}$ are the zeroes of the function
\begin{equation}
\label{Q[erf]}
\mathcal{Q}(x)=1+\erf\Big(\frac{x}{2\sqrt{2}}\Big)\;,
\end{equation}
or equivalently $\mathcal{Q}(x)=\mathrm{erfc}(-x/(2\sqrt{2}))$, whose Weierstrass factorization \cite{C1978.1} is then given by
\begin{equation}
\label{Q mu=0 Weierstrass}
\mathcal{Q}(x)=\rme^{\frac{x}{\sqrt{2\pi}}-\frac{x^{2}}{4\pi}}\prod_{j\in\mathbb{Z}^{*}}\Big[\Big(1-\frac{x}{w_{j}(0)}\Big)\rme^{\frac{x}{w_{j}(0)}+\frac{x^{2}}{2w_{j}(0)^{2}}}\Big]\;.
\end{equation}
The first $w_{j}(0)$ are plotted from (\ref{Q[erf]}) in figure \ref{fig w0}.

\begin{figure}
  \begin{center}
    \includegraphics[width=60mm]{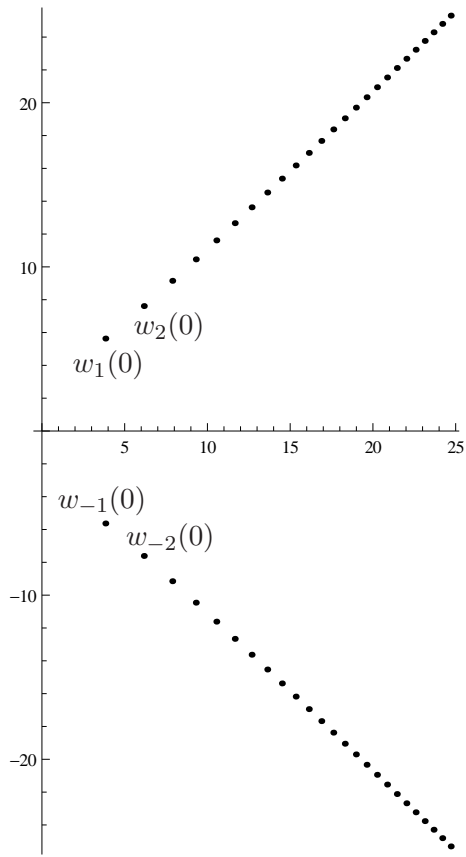}
    \begin{picture}(0,0)
      \put(-53,64){\small$w_{1}(0)$}
      \put(-45,69){\small$w_{2}(0)$}
      \put(-55,46){\small$w_{-1}(0)$}
      \put(-46,41){\small$w_{-2}(0)$}
    \end{picture}
  \end{center}
  \caption{Fifty first zeroes $w_{j}(0)$ of $\mathcal{Q}(x)=1+\erf(\frac{x}{2\sqrt{2}})$ in the complex plane.}
  \label{fig w0}
\end{figure}

The claim (\ref{Q[erf]}) can be checked by showing that the zeroes of (\ref{Q[erf]}) verify the edge Bethe equations (\ref{BE[w0]}). Starting with the equality $\mathcal{Q}''(x)+\frac{x}{4}\,\mathcal{Q}'(x)=0$ verified by (\ref{Q[erf]}), the residue at $x=w_{j}$ of $(\mathcal{Q}''(x)+\frac{x}{4}\,\mathcal{Q}'(x))/\mathcal{Q}(x)$ is then equal to $0$. Using (\ref{Q mu=0 Weierstrass}), this residue can also be computed in terms of the $w_{j}(0)$. It leads to
\begin{equation}
\fl\hspace{5mm}
\frac{w_{j}(0)}{4}-\frac{w_{j}(0)}{\pi}+\frac{2}{\sqrt{2\pi}}+\frac{4}{w_{j}(0)}+\sum_{k\in\mathbb{Z}^{*} \atop k\neq j}\Big(\frac{2w_{j}(0)}{w_{k}(0)^{2}}+\frac{2}{w_{k}(0)}+\frac{2}{w_{j}(0)-w_{k}(0)}\Big)=0\;.
\end{equation}
The identities
\begin{eqnarray}
\label{sum 1/w}
&& \sum_{j\in\mathbb{Z}^{*}}\Big(\frac{1}{w_{j}(0)}-\frac{1}{4\sqrt{\rmi\pi j}}\Big)=-\frac{1}{\sqrt{2\pi}}-\frac{\zeta(1/2)}{2\sqrt{2\pi}}\\
\label{sum 1/w2}
&& \sum_{j\in\mathbb{Z}^{*}}\frac{1}{w_{j}(0)^{2}}=-\frac{3}{16}+\frac{1}{2\pi}\;,
\end{eqnarray}
which are proven by residues in \ref{appendix sum w0} from (\ref{Q[erf]}) with a careful treatment of contours at infinity, finally lead to (\ref{BE[w0]}).

Some care is needed when dealing with infinite sums here. In particular, we note that a naive expansion at large $x$ of $\mathcal{Q}'(x)/\mathcal{Q}(x)$ from (\ref{Q mu=0 Weierstrass}) gives the incorrect coefficient $-1/4$ instead of $-3/16$ in (\ref{sum 1/w2}) due to the impossibility to exchange the sum over $j\in\mathbb{Z}^{*}$ with the limit $x\to\infty$. In the residue calculation of \ref{appendix sum 1/w0*w0}, an extra term $-1/16$ contributed by contours at infinity gives the correct result (\ref{sum 1/w2}).

The interpretation of the $w_{j}(0)$ as zeroes of a simple function is reminiscent of the small coupling limit of the Lieb-Liniger $\delta$-Bose gas \cite{LL1963.1} with $N$ particles on a circle, for which the Bethe roots for the ground state converge to zeroes of the $N$-th Hermite polynomial $H_{N}$ \cite{GC2014.1}. A simple proof of that relies also on the second order linear differential equation satisfied by $H_{N}$.

Additionally, the zeroes of $\mathcal{Q}$ also appear when studying the large $N$ limit of the truncated exponential function $\exp_{N}(z)=\sum_{j=0}^{N}(Nz)^{j}/j!$, whose zeroes accumulate on a closed curve called the Szeg\"o curve \cite{S1922.1}, with the same kind of singularity as the curve of figure \ref{fig curve y} on which the Bethe roots $y_{j}$ accumulate. It turns out that the zeroes of $\exp_{N}$ close to the singularity also converge \cite{NR1972.1} to the zeroes of $1+\erf$ after rescaling by a factor $\sqrt{N}$. This result was extended recently \cite{V2015.1} to any entire function with reasonable growth at infinity.

The evidence presented here for (\ref{Q[erf]}) is entirely non-constructive, as we postulate (\ref{Q[erf]}) and then derive the edge Bethe equations (\ref{BE[w0]}). Our original ``derivation'' of (\ref{Q[erf]}) was based on the assumption $w_{j}(0)\simeq4\sqrt{\rmi\pi|j|+d_{\sgn(j)}}$ for large $|j|$. Then, tedious Euler-Maclaurin asymptotics lead us to a second order differential equation for $\mathcal{Q}$, whose solution reduces to (\ref{Q[erf]}) when $d_{+}-d_{-}=1/2$. The initial assumption for the large $|j|$ asymptotics of $w_{j}(0)$ is however invalidated by the final result (\ref{Q[erf]}), from which one finds $(w_{j}(0)/4)^{2}\simeq\rmi\pi|j|-\frac{\log(8\rmi\pi^{2}|j|)}{4}+\mathcal{O}(\frac{\log|j|}{|j|})$ when $|j|\to\infty$, with higher powers of $\log|j|$ in sub-leading terms.
\end{subsection}

\begin{subsection}{Unproven assumptions of section \ref{section direct}}
The whole Bethe ansatz approach to the spectrum of $M$ is based on the unproven assumption that any eigenstate of $M$ corresponds to a solution of the Bethe equations (\ref{BE[y]}), see however \cite{LSA1995.1,LSA1997.1,BDS2015.1} for progress about such completeness issues. Since we study the gap, only the existence of the corresponding solution to the Bethe equations is in fact needed, but proving it is unlikely to be simpler than proving the whole completeness. An additional difficult issue is showing that the choice of the numbers $n_{j}$ below (\ref{f[yj]}) correctly identifies the solution of the logarithmic Bethe equations (\ref{LBE[y]}) corresponding to the gap. Another implicit assumption is the existence, as in figure \ref{fig branch cuts f}, of a contour from $-1+\rmi0^{+}$ to $-1+\rmi0^{-}$ that does not cross any branch cut of the function $f$, at least for a large enough value of the system size $L$.

Under the previous assumptions, the Euler-Maclaurin calculations leading to the edge Bethe equations (\ref{BE[w]}), (\ref{g(w)}) can presumably be made rigorous without too much effort. Additionally, the limit $\mu\to0$ of the edge Bethe equations (\ref{BE[w0]}) uses the unproven assumption that $w_{0}(\mu)$ goes to infinity when $\mu\to0$, while the $w_{j}(\mu)$, $j\in\mathbb{Z}^{*}$ converge to distinct finite values $w_{j}(0)$. Finally, our derivation of (\ref{Q[erf]}) requires that the edge Bethe equations (\ref{BE[w0]}) have a unique solution.
\end{subsection}
\end{section}

\begin{section}{Functional approach}
\label{section functional}
In this section, we introduce simple functional equations from which a systematic small $\mu$ expansion of the spectral gap of WASEP can be extracted. The starting point of the perturbative expansion is the generating function (\ref{Q[erf]}) of the edge Bethe roots obtained in the previous section in the limit $\mu\to0$. We emphasize that several important steps in our derivation of the functional equations are guessed from high precision extrapolation numerics.

\begin{subsection}{Baxter's equation}
We introduce Baxter's Q polynomial, that we call $\hat{Q}$ here \footnote{The polynomial $\hat{Q}$ should not be confused with the function $\mathcal{Q}$ defined as (\ref{Q[erf]}) in section \ref{section direct}.}, defined as
\begin{equation}
\label{Q[yj]}
\hat{Q}(y)=\prod_{j=1}^{N}(y-y_{j})\;,
\end{equation}
and whose zeroes are the Bethe roots $y_{j}$ of a finite system. Introducing the polynomial $\hat{R}(y)=(1-y)^{L}\hat{Q}(qy)+q^{N}(1-qy)^{L}\hat{Q}(q^{-1}y)$, the Bethe equations (\ref{BE[y]}) can then be written as $\hat{R}(y_{j})=0$, $j=1,\ldots,N$, implying that $\hat{Q}(y)$ divides $\hat{R}(y)$. The ratio $\hat{T}(y)$ turns out to be equal to the eigenvalue of the transfer matrix built in the algebraic Bethe ansatz framework. One has
\begin{equation}
\label{ThQh eq}
\hat{T}(y)\hat{Q}(y)=(1-y)^{L}\hat{Q}(qy)+q^{N}(1-qy)^{L}\hat{Q}(q^{-1}y)\;.
\end{equation}
This functional equation alternatively follows from an operator identity \cite{B1982.1,LP2014.1} between the transfer matrix and an operator version of $\hat{Q}$. A modified version of (\ref{ThQh eq}) incorporating a fugacity conjugate to the current was used for ASEP with periodic \cite{PM2008.1,PM2009.1,P2010.1,S2011.1} and open \cite{LP2014.1} boundary conditions in order to compute large deviations of the current.

\begin{subsubsection}{Large \texorpdfstring{$L$}{L} limit of \texorpdfstring{$\hat{Q}$}{Q hat}}
We focus again on the half-filled system $\rho=1/2$. We are interested in $\hat{Q}(y)$ with $y+1\sim1/\sqrt{L}$ compatible with the scaling of the edge Bethe roots, and write $y=-1+x/\sqrt{L}$.

We want the large $L$ asymptotics of $\hat{Q}(-1+x/\sqrt{L})$. The terms corresponding to $j$ at a distance at most $M$ from the edges $1$ and $N$ in (\ref{Q[yj]}) contribute $-(M+\frac{1}{2})\log L+\sum_{j=-M}^{M}\log(x-w_{j})$ to $\log\hat{Q}(y)$ for finite $M$. However, unlike in section \ref{subsection f edge}, we observe that it is not possible to obtain the asymptotics of the bulk of the sum by simply replacing $y_{j}=f^{-1}(2\rmi\pi n_{j}/L)$ by $f_{0}^{-1}(2\rmi\pi n_{j}/L)$ for values of $j$ far from the boundaries: we would also need for this the correction of order $L^{-1}$, which is not easily accessible. Hence, we did not manage to properly derive the large $L$ limit of $\hat{Q}$ in terms of the reduced Bethe roots $w_{j}$.

The difficulty can be bypassed with the help of the high precision extrapolation methods described in \cite{P2016.2}: extrapolating $\hat{Q}(y)$ on $L$ from Bethe ansatz numerics, we were able to guess how to properly regularize $\hat{Q}(-1+x/\sqrt{L})$ in the large $L$ limit. We claim that $(-1)^{N}\rme^{\frac{x\sqrt{L}}{2}}\hat{Q}(-1+x/\sqrt{L})$ is finite when $L\to\infty$, and define \footnote{The function $Q_{\mu}$ should not be confused with the function $\mathcal{Q}$ defined as (\ref{Q[erf]}) in section \ref{section direct}; in the limit $\mu\to0$, the function $Q_{\mu}$ is however proportional to $\mathcal{Q}$, as shown below in equation (\ref{Q mu 0}).}
\begin{equation}
\label{Q[Qh]}
Q_{\mu}(x)=\lim_{L\to\infty}\rme^{\frac{x^{2}}{4}}(-1)^{N}\rme^{\frac{x\sqrt{L}}{2}}\hat{Q}(-1+x/\sqrt{L})\;.
\end{equation}
The factor $\rme^{\frac{x^{2}}{4}}$ is for later convenience. Unlike with $\hat{Q}$, we choose to keep track explicitly of the dependency in $\mu$. From (\ref{ThQh eq}), the quantity $2^{-L}\rme^{\frac{(x+\mu)\sqrt{L}}{2}}\hat{T}(-1+x/\sqrt{L})$ also has a finite limit when $L\to\infty$, and we define
\begin{equation}
\label{T[Th]}
T_{\mu}(x)=\lim_{L\to\infty}\rme^{\frac{x^{2}+\mu^{2}}{4}}2^{-L}\rme^{\frac{(x+\mu)\sqrt{L}}{2}}\hat{T}(-1+x/\sqrt{L})\;.
\end{equation}
Baxter's functional equation (\ref{ThQh eq}) then reduces for $T_{\mu}$ and $Q_{\mu}$ to
\begin{equation}
\label{TQ eq}
T_{\mu}(x)Q_{\mu}(x)=\rme^{\frac{x^{2}}{8}}Q_{\mu}(x+\mu)+\rme^{\frac{(x+\mu)^{2}}{8}}Q_{\mu}(x-\mu)\;.
\end{equation}
\end{subsubsection}

\begin{subsubsection}{Limit \texorpdfstring{$x\to-\infty$}{x->infinity} of \texorpdfstring{$Q_{\mu}(x)$}{Q_mu(x)}}
Momentum, energy and higher local conserved charges of quantum integrable models are obtained in the Bethe ansatz framework by taking logarithmic derivatives of the eigenvalue of the transfer matrix $\hat{T}(y)$ at a specific value for $y$, $y=1$ here with our notations. This value lies outside the scaling regime $y=-1+x/\sqrt{L}$, and hence should correspond to $|x|\to\infty$. Since the edge Bethe roots $w_{j}(\mu)$ behave as $w_{j}\simeq4\sqrt{\rmi\pi j}$ when $|j|\to\infty$, as observed from numerics, we expect a different large $|x|$ behaviour for $Q_{\mu}(x)$ depending on whether $-\frac{\pi}{4}<\arg(x)<\frac{\pi}{4}$ or not. The point $y=1$ lies outside the closed curve on which the Bethe roots $y_{j}$ accumulate, see figures \ref{fig branch cuts f} and \ref{fig curve y}, and thus corresponds to $|x|\to\infty$ with $-\frac{3\pi}{4}<\arg(-x)<\frac{3\pi}{4}$.

The large $|x|$ behaviour of $Q_{\mu}(x)$ can be explored numerically from double extrapolation on $L$ and $x$. We conjecture the expansion
\begin{equation}
\label{Q(-infty)}
Q_{\mu}(x)\underset{|x|\to\infty}{\simeq}1+\frac{8\rmi\pi}{\mu x}+\frac{c(\mu)}{x^{2}}
\end{equation}
in the region $-\frac{3\pi}{4}<\arg(-x)<\frac{3\pi}{4}$. The coefficient of the $x^{-1}$ term corresponds to the momentum of the gap eigenstate. The coefficient $c(\mu)$ of the $x^{-2}$ term will be shortly related to the eigenvalue.
\end{subsubsection}

\begin{subsubsection}{Large \texorpdfstring{$L$}{L} limit of the eigenvalue}
The eigenvalue (\ref{E[y]}) corresponding to the gap eigenstate is
\begin{eqnarray}
\label{E[Q]}
&& E_{1}=(1-q)\Bigg(\frac{\hat{Q}'(1)}{\hat{Q}(1)}-q^{-1}\frac{\hat{Q}'(q^{-1})}{\hat{Q}(q^{-1})}\Bigg)
\\
\label{E[T]}
&&\hspace{4.7mm} =-qL-(1-q)\,\frac{\hat{T}'(1)}{\hat{T}(1)}\;.
\end{eqnarray}
This expression involves evaluating $\hat{Q}(y)$ (or $\hat{T}(y)$) around $y=1$, outside the scaling regime for $Q_{\mu}$. Using nonetheless (\ref{Q[Qh]}) naively leads, in conjunction with (\ref{Q(-infty)}), to
\begin{equation}
\label{E[c] wrong}
2\mu^{2}+\frac{2\mu^{3}}{\sqrt{L}}+\frac{2\mu^{4}}{L}+\frac{2\mu^{5}}{L^{3/2}}+\frac{2\mu^{6}-4\pi^{2}-(-\frac{3\rmi\pi}{2}+\frac{c(\mu)}{8})\mu^{2}}{L^{2}}
\end{equation}
for the right hand side of (\ref{E[Q]}), with $c(\mu)$ the coefficient of $x^{-2}$ in the large $x$ expansion (\ref{Q(-infty)}) of $Q_{\mu}(x)$. This can not be correct since we expect $E_{1}\sim L^{-2}$ for large $L$. The problem comes again from the bulk of the Bethe roots $y_{j}$, which should however only cancel the divergent contributions to $L^{2}E_{1}$ in (\ref{E[c] wrong}), and modify in a hopefully simple way the constant term. Non-trivial terms are only expected to be contributed by the edge Bethe roots. We conjecture from high precision numerical extrapolation that one has in fact $E_{1}=e_{1}(\mu)/L^{2}+\mathcal{O}(L^{-5/2})$ with
\begin{equation}
\label{e1[c]}
e_{1}(\mu)=-4\pi^{2}-\Big(\frac{\rmi\pi}{2}+\frac{c(\mu)}{8}\Big)\mu^{2}\;,
\end{equation}
which allows to compute the eigenvalue knowing $Q_{\mu}$.
\end{subsubsection}

\begin{subsubsection}{Small \texorpdfstring{$\mu$}{mu} limit}
From (\ref{Q[erf]}), we expect that $Q_{\mu}(x)$ should be divisible by $1+\erf(\frac{x}{2\sqrt{2}})$ when $\mu\to0$. The ratio can be guessed again using high precision numerics for $\hat{Q}(y)$ extrapolated on $L$ and $\mu$. We conjecture
\begin{equation}
\label{Q mu 0}
Q_{\mu}(x)\underset{\mu\to0}{\simeq}-\rmi(2\pi)^{3/2}\rme^{\frac{x^{2}}{8}}\Big(1+\erf\Big(\frac{x}{2\sqrt{2}}\Big)\Big)\mu^{-1}\;.
\end{equation}
From (\ref{TQ eq}), the small $\mu$ limit of $T_{\mu}(y)$ is then equal to
\begin{equation}
\label{T mu 0}
T_{0}(x)=2\,\rme^{\frac{x^{2}}{8}}\;.
\end{equation}
\end{subsubsection}

\begin{figure}
  \begin{center}
    \includegraphics[width=75mm]{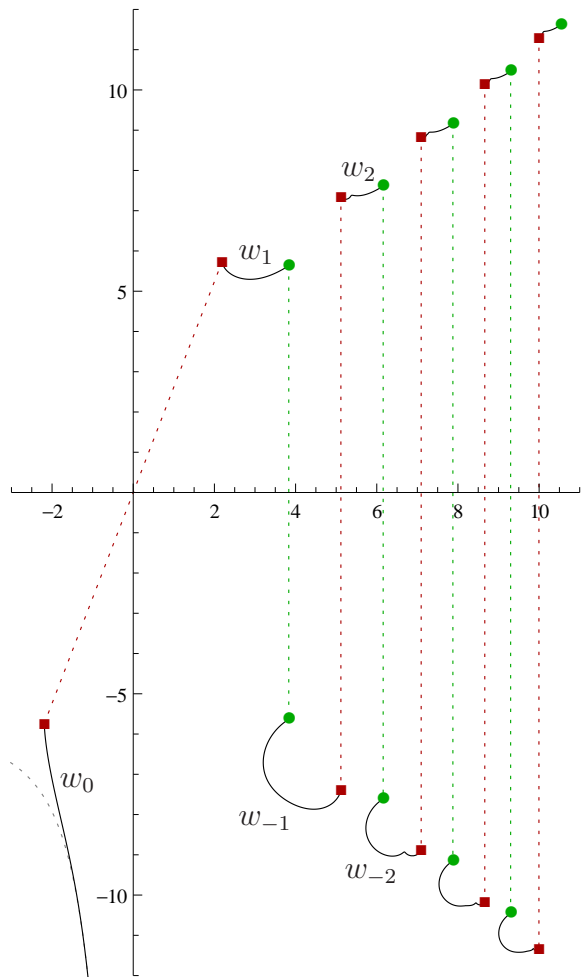}
    \begin{picture}(0,0)
      \put(-46.5,95){$w_{1}$}
      \put(-33,106){$w_{2}$}
      \put(-46.5,20){$w_{-1}$}
      \put(-32.5,13){$w_{-2}$}
      \put(-70,25){$w_{0}$}
    \end{picture}
  \end{center}
  \caption{Edge Bethe roots $w_{j}(\mu)$, $j=-4,\ldots,5$ plotted in the complex plane. The green dots correspond to the symmetric limit $\mu=0$, and the red squares to the totally asymmetric limit $\mu\to\infty$. The solid black curves represent finite asymmetry $\mu>0$. The dotted green lines illustrate that $w_{j}(0)$ and $w_{-j}(0)$, $j\in\mathbb{N}^{*}$ are related by complex conjugation. The dotted red lines illustrate that $w_{j+1}(\infty)$ and $w_{-j}(\infty)$, $j\in\mathbb{N}^{*}$ are related by complex conjugation, and that $w_{0}(\infty)=-w_{1}(\infty)$.}
  \label{fig wj(mu)}
\end{figure}

\begin{subsubsection}{Large \texorpdfstring{$\mu$}{mu} limit}
The limit of large asymmetry $\mu\to\infty$ can be computed from TASEP ($q=0$). Slightly tedious Euler-Maclaurin calculations similar to the ones in \cite{P2015.2} lead to
\begin{equation}
\label{Q mu infinity}
\fl\hspace{10mm}
Q_{\infty}(x)=\frac{\sqrt{\nu_{1}+\frac{x^{2}}{8}-\rmi\pi}}{\sqrt{\nu_{1}+\frac{x^{2}}{8}+\rmi\pi}}\,\sqrt{1+\rme^{\nu_{1}+\frac{x^{2}}{8}}}\,\exp\Bigg(\int_{\nu_{1}}^{\nu_{1}+\frac{x^{2}}{8}}\rmd v\,\frac{\sgn(\Re\,x)\,\varphi_{1}'(v)}{\sqrt{2}\sqrt{\nu_{1}+\frac{x^{2}}{8}-v}}\Bigg)\;,
\end{equation}
where $\varphi_{1}(v)=-(2\pi)^{-1/2}\Li_{3/2}(-\rme^{v})-\sqrt{2\rmi\pi-2v}-\sqrt{-2\rmi\pi-2v}$, with $\Li_{3/2}$ the (analytic continuation of the) polylogarithm and $\nu_{1}$ the solution of $\varphi_{1}(\nu_{1})=0$, $\nu_{1}\approx3.5156583$. The zeroes of $Q_{\infty}$ are $w_{j}(\infty)=4\sqrt{\rmi\pi j-(\nu_{1}+\rmi\pi)/2}$, $j\in\mathbb{Z}^{*}$, and $w_{0}(\infty)=-w_{1}(\infty)$. They are plotted in figure \ref{fig wj(mu)}.
\end{subsubsection}

\end{subsection}

\begin{subsection}{Quantum Wronskian}
It is convenient to supplement Baxter's function $Q_{\mu}$ with another one, $P_{\mu}$, obtained by exchanging particles and empty sites in the Bethe ansatz formalism. A symmetry relation discussed below between $P_{\mu}$ and $Q_{\mu}$ then leads to a systematic expansion of the gap eigenvalue $E_{1}$ in the asymmetry $\mu$.

\begin{subsubsection}{Baxter's equation ``beyond the equator''}
Exchanging particles and empty sites in the Bethe ansatz leads to an alternative set of Bethe equations for $L-N$ Bethe roots ``beyond the equator'' $\tilde{y}_{j}$, $j=1,\ldots,L-N$ \cite{PS1999.1}. The corresponding Baxter's polynomial is $\hat{P}(y)=\prod_{j=1}^{L-N}(y-\tilde{y}_{j})$. The polynomial $\hat{P}$ verifies \cite{PS1999.1,P2010.1} a functional equation analogous to (\ref{ThQh eq}):
\begin{equation}
\label{ThPh eq}
\hat{T}(y)\hat{P}(y)=q^{N}(1-y)^{L}\hat{P}(qy)+(1-qy)^{L}\hat{P}(q^{-1}y)\;,
\end{equation}
with the same polynomial $\hat{T}$ of degree $L$ as before. The proper way to extract the large $L$ limit of $\hat{P}(y)$ for WASEP $q=1-\mu/\sqrt{L}$ with $y=-1+x/\sqrt{L}$ can be guessed from extrapolated numerics as before. We focus again on the half-filled case $\rho=1/2$ for simplicity, and define
\begin{equation}
\label{P[Ph]}
P_{\mu}(x)=\lim_{L\to\infty}\hat{P}(0)^{-1}\hat{P}(-1+x/\sqrt{L})\;.
\end{equation}
Then, (\ref{ThPh eq}) leads to Baxter's equation for $P_{\mu}$:
\begin{equation}
\label{TP eq}
T_{\mu}(x)P_{\mu}(x)=\rme^{\frac{x^{2}}{8}}P_{\mu}(x+\mu)+\rme^{\frac{(x+\mu)^{2}}{8}}P_{\mu}(x-\mu)\;,
\end{equation}
which is exactly the same as the equation (\ref{TQ eq}) for $Q_{\mu}$. The functions $P_{\mu}$ and $Q_{\mu}$ are however different, since their Wronskian (\ref{QP eq}) below is non-zero.
\end{subsubsection}

\begin{subsubsection}{Quantum Wronskian}
Assuming $\hat{T}$, $\hat{Q}$ and $\hat{P}$ are polynomials of respective degrees $L$, $N$ and $L-N$, simple manipulations \cite{PS1999.1,P2010.1} of both Baxter's equations (\ref{ThQh eq}), (\ref{ThPh eq}) lead to
\begin{eqnarray}
\label{QhPh eq}
&&(1-q^{N})\hat{Q}(0)\hat{P}(0)(1-y)^{L}=\hat{Q}(y)\hat{P}(q^{-1}y)-q^{N}\hat{Q}(q^{-1}y)\hat{P}(y)\\
\label{ThQhPh eq}
&&(1-q^{N})\hat{Q}(0)\hat{P}(0)\hat{T}(y)=\hat{Q}(qy)\hat{P}(q^{-1}y)-q^{2N}\hat{Q}(q^{-1}y)\hat{P}(qy)\;.
\end{eqnarray}
Equation (\ref{QhPh eq}), called the quantum Wronskian, is a discrete analogue of the Wronskian of a linear ordinary differential equation of second order.

Large $L$ asymptotics of all the pieces in (\ref{QhPh eq}), (\ref{ThQhPh eq}) are given above as (\ref{Q[Qh]}), (\ref{P[Ph]}) and (\ref{T[Th]}), except for $\hat{Q}(0)$ which is outside the scaling limit (\ref{Q[Qh]}). We define
\begin{equation}
\label{C[Qh]}
C_{\mu}=\lim_{L\to\infty}(-1)^{N}2^{L}(1-q^{N})\hat{Q}(0)\;.
\end{equation}
Then, (\ref{QhPh eq}) and (\ref{ThQhPh eq}) supplemented with (\ref{Q[Qh]}), (\ref{T[Th]}), (\ref{P[Ph]}) lead to the functional equations
\begin{eqnarray}
\label{QP eq}
&& C_{\mu}\,\rme^{\frac{x^{2}}{8}}=Q_{\mu}(x)P_{\mu}(x-\mu)-Q_{\mu}(x-\mu)P_{\mu}(x)\\
\label{TQP eq}
&& C_{\mu}\,T_{\mu}(x)=Q_{\mu}(x+\mu)P_{\mu}(x-\mu)-Q_{\mu}(x-\mu)P_{\mu}(x+\mu)\;.
\end{eqnarray}
In the following, we only make use of the Wronskian identity (\ref{QP eq}).
\end{subsubsection}

\begin{figure}
  \begin{center}
    \includegraphics[width=100mm]{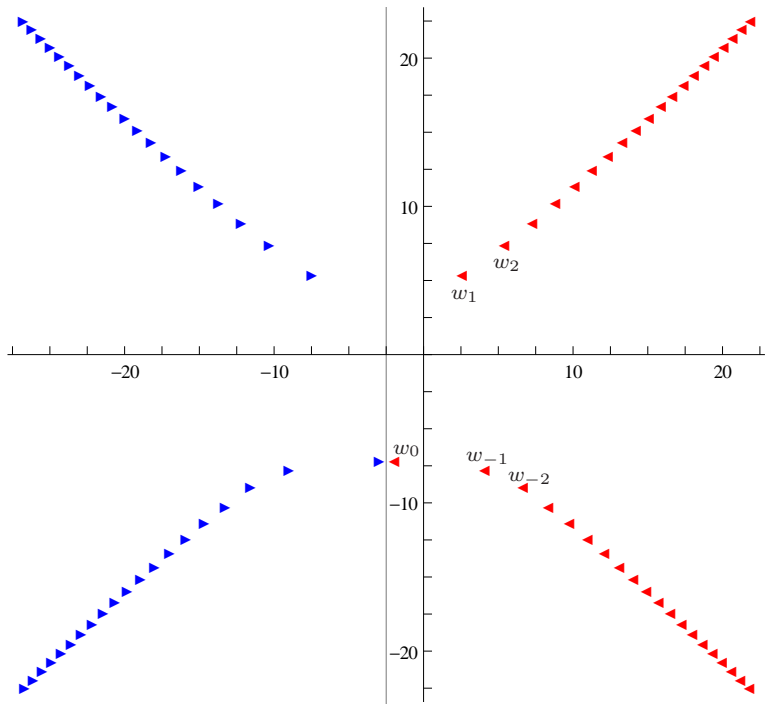}
    \begin{picture}(0,0)
      \put(-50.6,33){\scriptsize$w_{0}$}
      \put(-43,53.5){\scriptsize$w_{1}$}
      \put(-37.5,57.6){\scriptsize$w_{2}$}
      \put(-41,32){\scriptsize$w_{-1}$}
      \put(-35.5,29.4){\scriptsize$w_{-2}$}
    \end{picture}
  \end{center}
  \caption{First edge Bethe roots at $\mu=5$. The zeroes $w_{j}(\mu)$ of $Q_{\mu}$ are represented by the red triangles \textcolor{red}{\tiny$\blacktriangleleft$} on the right of the vertical gray line of abscissa $-\mu/2$. The zeroes $-\overline{w_{j}(\mu)}-\mu$ of $P_{\mu}$ are represented by the blue triangles \textcolor{blue}{\tiny$\blacktriangleright$} on the left of the vertical gray line of abscissa $-\mu/2$.}
  \label{fig wj QP}
\end{figure}

\begin{subsubsection}{Symmetry relation between \texorpdfstring{$P_{\mu}$}{P_mu} and \texorpdfstring{$Q_{\mu}$}{Q_mu}}
The functions $Q_{\mu}$ and $P_{\mu}$ are distinct solutions of the exact same Baxter's equation (\ref{TQ eq}), (\ref{TP eq}). The form of the two exponentials in the right hand side of (\ref{TQ eq}), (\ref{TP eq}) suggests that $Q_{\mu}$ and $P_{\mu}$ might be exchanged by the transformation $x\to-x-\mu$. This would however change the sign of the momentum in the large $|x|$ expansion (\ref{Q(-infty)}). A better guess is the replacement $x\to-\overline{x}-\mu$, where $\overline{\,\cdot\,}$ is complex conjugation, followed by a global complex conjugation. Numerics confirm the symmetry relations
\begin{equation}
\label{symmetry T}
T_{\mu}(x)=\overline{T_{\mu}(-\overline{x}-\mu)}
\end{equation}
and
\begin{equation}
\label{symmetry QP}
P_{\mu}(x)=\overline{Q_{\mu}(-\overline{x}-\mu)}\;,
\end{equation}
which correspond to a reflection of the complex plane with respect to the vertical line of abscissa $-\mu/2$, see figure \ref{fig wj QP}. We observe numerically that the zeroes of $Q_{\mu}$ stay on the right of the vertical line for all $\mu>0$, and hence the zeroes of $P_{\mu}$ stay on the left of that line. When $\mu\to0$, $w_{0}(\mu)\simeq-8\rmi\pi/\mu$, which is compatible with (\ref{Q(-infty)}). We also observe numerically that $\Re\,w_{0}(\mu)+\mu/2$ converges to $0$ exponentially fast at small $\mu$, so that the corresponding zero of $P_{\mu}$ is exponentially close to $w_{0}(\mu)$.

The relation (\ref{symmetry QP}) fixes in particular the small $\mu$ behaviour
\begin{equation}
\label{P_0}
P_{\mu}(x)\underset{\mu\to0}{\simeq}\rmi(2\pi)^{3/2}\rme^{\frac{x^{2}}{8}}\Big(1-\erf\Big(\frac{x}{2\sqrt{2}}\Big)\Big)\mu^{-1}
\end{equation}
and the large $|x|$ asymptotics
\begin{equation}
\label{P(infty)}
P_{\mu}(x)\underset{|x|\to\infty}{\simeq}1+\frac{8\rmi\pi}{\mu x}+\frac{c(\mu)}{x^{2}}
\end{equation}
in the region $-\frac{3\pi}{4}<\arg x<\frac{3\pi}{4}$ using $\overline{c(\mu)}-8\rmi\pi=c(\mu)$, which comes from the fact that $e_{1}(\mu)$ is real. The overlap of the domains of validity of the expansions (\ref{Q(-infty)}) and (\ref{P(infty)}) is $\frac{\pi}{4}<|\arg x|<\frac{3\pi}{4}$, which corresponds to the region $\Re\,x^{2}<0$ where the left hand side of (\ref{QP eq}) vanishes, suggesting that the large $|x|$ expansions of $Q_{\mu}$ and $P_{\mu}$ are in fact equal at any order in $x$ in the intersection of their respective domains of validity.
\end{subsubsection}

\begin{subsubsection}{Reduction of the Wronskian}
We conjecture from numerics that $\rme^{-\frac{x^{2}}{8}}T_{\mu}(x)$ has an expansion near $\mu=0$ of the form $\sum_{k=0}^{\infty}\tau_{k}(x)\mu^{k}$ where the $\tau_{k}$ are polynomials. Using (\ref{TQ eq}), (\ref{Q(-infty)}) and (\ref{Q mu 0}), this implies that $Q_{\mu}$ can be written as
\begin{equation}
\label{Q[alpha,beta]}
Q_{\mu}(x)=\alpha_{\mu}(x)\,\rme^{\frac{x^{2}}{8}}\,\Big(1+\erf\Big(\frac{x}{2\sqrt{2}}\Big)\Big)+\beta_{\mu}(x)\;,
\end{equation}
where $\alpha_{\mu}(x)$ and $\beta_{\mu}(x)$ are polynomials in $x$ at each order in $\mu$. Similarly,
\begin{equation}
\label{P[gamma,delta]}
P_{\mu}(x)=\gamma_{\mu}(x)\,\rme^{\frac{x^{2}}{8}}\,\Big(1-\erf\Big(\frac{x}{2\sqrt{2}}\Big)\Big)+\delta_{\mu}(x)\;.
\end{equation}
Multiplying (\ref{QP eq}) by
\begin{equation}
\frac{1}{1+\erf(\frac{x}{2\sqrt{2}})}\,\frac{1}{1-\erf(\frac{x}{2\sqrt{2}})}=\frac{1/2}{1+\erf(\frac{x}{2\sqrt{2}})}+\frac{1/2}{1-\erf(\frac{x}{2\sqrt{2}})}\;,
\end{equation}
and writing $\erf(\frac{x-\mu}{2\sqrt{2}})=\erf(\frac{x}{2\sqrt{2}})+\epsilon_{\mu}(x)$, one finds an expression of the form
\begin{equation}
\label{expr[U,V,W]}
U+\frac{V}{1+\erf(\frac{x}{2\sqrt{2}})}+\frac{W}{1-\erf(\frac{x}{2\sqrt{2}})}=0\;,
\end{equation}
where $U$, $V$, $W$ do not contain error functions, except implicitly in $\epsilon_{\mu}(x)$. The expressions $U$, $V$ and $W$ must be identically equal to zero. In particular, $U=0$ implies
\begin{equation}
\label{U=0}
\frac{\alpha_{\mu}(x)}{\gamma_{\mu}(x)}=\frac{\alpha_{\mu}(x-\mu)}{\gamma_{\mu}(x-\mu)}\;,
\end{equation}
while $V-W=0$ in conjunction with (\ref{U=0}) leads to
\begin{equation}
\label{V-W=0}
\rme^{-\frac{x^{2}}{8}}\Big(\frac{\beta_{\mu}(x)}{\alpha_{\mu}(x)}+\frac{\delta_{\mu}(x)}{\gamma_{\mu}(x)}\Big)=\rme^{-\frac{(x-\mu)^{2}}{8}}\Big(\frac{\beta_{\mu}(x-\mu)}{\alpha_{\mu}(x-\mu)}+\frac{\delta_{\mu}(x-\mu)}{\gamma_{\mu}(x-\mu)}\Big)\;.
\end{equation}
Since $\alpha_{\mu}$, $\beta_{\mu}$, $\gamma_{\mu}$ and $\delta_{\mu}$ are polynomials at each order in $\mu$, periodicity implies that $\alpha_{\mu}(x)/\gamma_{\mu}(x)$ is independent of $x$ and that $\beta_{\mu}(x)/\alpha_{\mu}(x)+\delta_{\mu}(x)/\gamma_{\mu}(x)=0$. Moreover, numerics indicate that the ratio $\alpha_{\mu}(x)/\gamma_{\mu}(x)$ is independent of $\mu$, and equal to $-1$ from (\ref{Q mu 0}) and (\ref{P_0}). Thus
\begin{equation}
\label{gamma[alpha] delta[beta]}
\gamma_{\mu}(x)=-\alpha_{\mu}(x)\qquad\text{and}\qquad\delta_{\mu}(x)=\beta_{\mu}(x)\;,
\end{equation}
and the last equation for the cancellation of (\ref{expr[U,V,W]}), $V+W=0$, becomes
\begin{equation}
\label{V+W=0}
\fl\hspace{5mm}\alpha_{\mu}(x)\Big(\beta_{\mu}(x-\mu)+\rme^{\frac{(x-\mu)^{2}}{8}}\epsilon_{\mu}(x)\alpha_{\mu}(x-\mu)\Big)-\rme^{\frac{(x-\mu)^{2}}{8}-\frac{x^{2}}{8}}\alpha_{\mu}(x-\mu)\beta_{\mu}(x)=\frac{C_{\mu}}{2}\;.
\end{equation}
This equation can be simplified further by noting that
\begin{eqnarray}
\label{alpha[Q,P]}
&& \rme^{\frac{x^{2}}{8}}\,\alpha_{\mu}(x)=\frac{Q_{\mu}(x)-P_{\mu}(x)}{2}\\
\label{beta[Q,P]}
&& \rme^{-\frac{x^{2}}{8}}\,\frac{\beta_{\mu}(x)}{\alpha_{\mu}(x)}+\erf\Big(\frac{x}{2\sqrt{2}}\Big)=\frac{Q_{\mu}(x)+P_{\mu}(x)}{Q_{\mu}(x)-P_{\mu}(x)}
\end{eqnarray}
are both anti-symmetric under reflection of $x$ with respect to the vertical axis of abscissa $-\mu/2$ using (\ref{symmetry QP}):
\begin{eqnarray}
\label{symmetry alpha}
&&\hspace{-15mm} \Big[\rme^{\frac{(x+\mu)^{2}}{8}}\overline{\alpha_{\mu}(-\overline{x}-\mu)}\Big]+\Big[\rme^{\frac{x^{2}}{8}}\alpha_{\mu}(x)\Big]=0\\
\label{symmetry beta}
&&\hspace{-15mm} \Bigg[-\erf\Big(\frac{x+\mu}{2\sqrt{2}}\Big)+\rme^{-\frac{(x+\mu)^{2}}{8}}\,\frac{\overline{\beta_{\mu}(-\overline{x}-\mu)}}{\overline{\alpha_{\mu}(-\overline{x}-\mu)}}\Bigg]+\Bigg[\erf\Big(\frac{x}{2\sqrt{2}}\Big)+\rme^{-\frac{x^{2}}{8}}\,\frac{\beta_{\mu}(x)}{\alpha_{\mu}(x)}\Bigg]=0\;.
\end{eqnarray}
With these relations, (\ref{V+W=0}) becomes the very simple functional equation
\begin{equation}
\label{alpha beta eq}
\alpha_{\mu}(x)\overline{\beta_{\mu}(-\overline{x})}+\overline{\alpha_{\mu}(-\overline{x})}\beta_{\mu}(x)=\frac{C_{\mu}}{2}\;.
\end{equation}
\end{subsubsection}
\end{subsection}

\begin{subsection}{Perturbative solution in \texorpdfstring{$\mu$}{mu}}
The functional equation (\ref{alpha beta eq}), supplemented with the symmetry relation (\ref{symmetry alpha}) and the large $|x|$, $-\frac{3\pi}{4}<\arg(-x)<\frac{3\pi}{4}$ asymptotics $Q_{\mu}(x)\simeq1+8\rmi\pi/\mu x$ with (\ref{Q[alpha,beta]}) can be solved systematically order by order in $\mu$ for $\alpha_{\mu}(x)$, $\beta_{\mu}(x)$ and $C_{\mu}$ under the assumption that both $\alpha_{\mu}(x)$ and $\beta_{\mu}(x)$ are polynomials in $x$ at each order in the expansion near $\mu=0$. In particular, $Q_{\mu}(x)\simeq1$ at large $|x|$ supplemented with the expansion
\begin{equation}
\rme^{\frac{x^{2}}{8}}\Big(1+\erf(\frac{x}{2\sqrt{2}})\Big)\simeq\frac{4}{\sqrt{2\pi}}\sum_{m=0}^{\infty}\frac{(-1)^{m-1}2^{m}(2m)!}{m!x^{2m+1}}
\end{equation}
in the sector $-\frac{3\pi}{4}<\arg(-x)<\frac{3\pi}{4}$ allows to compute $\beta_{\mu}(x)$ as
\begin{equation}
\label{beta[alpha]}
\beta_{\mu}(x)=1+\frac{4}{\sqrt{2\pi}}\Big[\alpha_{\mu}(x)\sum_{m=0}^{\infty}\frac{(-1)^{m}2^{m}(2m)!}{m!x^{2m+1}}\Big]_{+}\;,
\end{equation}
where $[\ldots]_{+}$ means keeping only non-negative powers of $x$ after expanding near $\mu=0$. Similarly, equating the coefficient of the term of order $x^{-1}$ in the large $|x|$ expansion of $Q_{\mu}(x)$ to $8\rmi\pi/\mu$ implies
\begin{equation}
\label{alpha(0)}
\sum_{m=0}^{\infty}\frac{(-1)^{m}2^{m}}{m!}\,\alpha_{\mu}^{(2m)}(0)=-\frac{2\rmi\pi\sqrt{2\pi}}{\mu}\;.
\end{equation}

We observe that the functional equations (\ref{alpha beta eq}) and (\ref{symmetry alpha}) then allow to solve order by order in $\mu$ the coefficient $C_{\mu}$ and the function $\alpha_{\mu}(x)$. Within two minutes on a single core of a personal computer, we obtain expansions up to order $20$ in $\mu$. The first orders are
\begin{eqnarray}
&& C_{\mu}\simeq
\frac{2(2\pi)^{5/2}}{\mu}
+\Big(\frac{(2\pi)^{5/2}}{24}-\frac{\sqrt{2\pi}}{2}\Big)
-\frac{\mu^{3}}{2(2\pi)^{3/2}}\\
&&\hspace{10mm} +\Big(\frac{(2\pi)^{5/2}}{193536}+\frac{\sqrt{2\pi}}{1536}+\frac{(2\pi)^{-3/2}}{96}-\frac{9(2\pi)^{-7/2}}{8}\Big)\mu^{5}+\ldots\nonumber
\end{eqnarray}
and
\begin{eqnarray}
\label{alpha(mu)}
&&\hspace{-25mm}
\frac{\rme^{\frac{(\rmi x-\frac{\mu}{2})^2}{8}}\,\alpha_{\mu}(\rmi x-\frac{\mu}{2})}{\rmi\,\rme^{-\frac{x^2}{8}}}
\simeq-\frac{(2\pi)^{3/2}}{\mu}
+\frac{\sqrt{2\pi}\,x}{4}
+\Big(-\frac{(2\pi)^{2}+24}{384}\,x^{2}+\frac{(2\pi)^{2}+24}{96}\Big)\frac{\mu}{\sqrt{2\pi}}\\
&&\hspace{31mm}
+\Big(\frac{(2\pi)^{2}+24}{1536}\,x^{3}-\frac{5(2\pi)^{2}+24}{384}\,x\Big)\frac{\mu^{2}}{(2\pi)^{3/2}}+\ldots\nonumber
\end{eqnarray}
The function $p_{\mu}(x)=\frac{\rmi\mu}{8\pi^{2}}\,\rme^{\frac{(\rmi x-\mu/2)^2}{8}}\,\alpha_{\mu}(\rmi x-\mu/2)$ converges when $\mu\to0$ to the centred Gaussian with variance $2$. We conjecture from the beginning of the small $\mu$ expansion of $\alpha_{\mu}$ that $p_{\mu}$ is a properly normalized probability density function for all $\mu$. Numerics at finite $\mu$ seem to confirm positivity. We also observe the appearance of Bernoulli numbers in higher order terms of the expansion (\ref{alpha(mu)}), which allows to guess the first terms of the resummation at $x\sim\mu^{-1}$, $\frac{\alpha_{\mu}(4z/\mu)}{\sqrt{2\pi}}\simeq\frac{4\pi^{2}z\mu^{-1}}{(z+2\rmi\pi)(1-\rme^{-z})}$ with poles at $z=2\rmi\pi m$, $m\in\mathbb{Z}^{*}$.

The eigenvalue can finally be computed using (\ref{beta[alpha]}), (\ref{alpha(mu)}) and (\ref{e1[c]}) with $c(\mu)$ given in (\ref{Q(-infty)}). We obtain
\begin{eqnarray}
\label{e1(mu)}
&&\fl
e_{1}(\mu)=
-4\pi^2
-\frac{\mu^2}{2}
+\Big(-\frac{1}{16\pi^2}+\frac{1}{96}\Big)\mu^4
+\Big(-\frac{7}{256\pi^4}+\frac{1}{384\pi^2}-\frac{1}{11520}\Big)\mu^6\nonumber\\
&&\fl\hspace{15mm}
+\Big(-\frac{77}{4096\pi^6}+\frac{7}{4096\pi^4}+\frac{1}{30720\pi^2}+\frac{11}{3870720}\Big)\mu^8\nonumber\\
&&\fl\hspace{15mm}
+\Big(-\frac{1093}{65536\pi^8}+\frac{77}{49152\pi^6}+\frac{3}{163840\pi^4}-\frac{11}{7741440\pi^2}-\frac{23}{185794560}\Big)\mu^{10}\nonumber\\
&&\fl\hspace{15mm}
+\Big(-\frac{18447}{1048576\pi^{10}}+\frac{5465}{3145728\pi^8}+\frac{79}{9437184\pi^6}\\
&&\fl\hspace{52mm}
-\frac{29}{49545216\pi^4}+\frac{23}{297271296\pi^2}+\frac{631}{98099527680}\Big)\mu^{12}\nonumber\\
&&\fl\hspace{15mm}
+\Big(-\frac{354819}{16777216\pi^{12}}+\frac{18447}{8388608\pi^{10}}-\frac{171}{83886080\pi^8}-\frac{235}{528482304\pi^6}\nonumber\\
&&\fl\hspace{35mm}
+\frac{137}{9909043200\pi^4}-\frac{631}{130799370240\pi^2}-\frac{150851}{396758089728000}\Big)\mu^{14}\nonumber\\
&&\fl\hspace{15mm}
+\Big(-\frac{7586889}{268435456\pi^{14}}+\frac{827911}{268435456\pi^{12}}-\frac{301}{16777216\pi^{10}}\nonumber\\
&&\fl\hspace{24mm}
-\frac{5929}{12079595520\pi^8}+\frac{227}{45298483200\pi^6}-\frac{67}{747424972800\pi^4}\nonumber\\
&&\fl\hspace{59mm}
+\frac{150851}{453437816832000\pi^2}+\frac{203473}{8161880702976000}\Big)\mu^{16}\nonumber\\
&&\fl\hspace{15mm}
+\Big(-\frac{177503401}{4294967296\pi^{16}}+\frac{2528963}{536870912\pi^{14}}-\frac{762311}{16106127360\pi^{12}}\nonumber\\
&&\fl\hspace{23mm}
-\frac{2633}{4227858432\pi^{10}}+\frac{6749}{1449551462400\pi^8}+\frac{31}{217998950400\pi^6}\nonumber\\
&&\fl\hspace{15mm}
-\frac{271147}{7141645615104000\pi^4}-\frac{203473}{8161880702976000\pi^2}-\frac{15417901}{8633456032481280000}\Big)\mu^{18}\nonumber\\
&&\fl\hspace{15mm}
+O(\mu^{20})\;.\nonumber
\end{eqnarray}
We recover the expressions guessed up to order $10$ in $\mu$ from numerics in \cite{P2016.2}. The coefficient of $\mu^{12}$ matches perfectly the numerical value obtained there (within all $35$ digits given, except the last one due to rounding).
\end{subsection}

\begin{subsection}{Unproven identities of section \ref{section functional}}
The derivation of the perturbative expansion (\ref{e1(mu)}) in this section is full of gaps, that are only overcome using numerics. We summarize them here: (i) existence of the (non-zero) large $L$ limits (\ref{Q[Qh]}), (\ref{P[Ph]}), (\ref{C[Qh]}) of $\hat{Q}(y)$, $\hat{P}(y)$ and $\hat{Q}(0)$. (ii) large $|x|$ expansion (\ref{Q(-infty)}) of $Q_{\mu}(x)$ and its relation (\ref{e1[c]}) to the gap eigenvalue $e_{1}(\mu)$. (iii) small $\mu$ limit of $Q_{\mu}(x)$ in (\ref{Q mu 0}) and appearance of polynomials in the small $\mu$ expansion of $T_{\mu}(x)$ above (\ref{Q[alpha,beta]}). (iv) symmetry relations (\ref{symmetry T}), (\ref{symmetry QP}). (v) complete derivation of (\ref{gamma[alpha] delta[beta]}) from (\ref{Q[alpha,beta]}), (\ref{P[gamma,delta]}).
\end{subsection}

\begin{subsection}{Higher excited states}
Higher excited states with an eigenvalue $E\simeq e(\mu)/L^{2}$ are needed in order to fully describe the relaxation in finite volume for the universal model labelled by $\mu$ on the renormalization group flow connecting the Edwards-Wilkinson and the KPZ fixed point. An important goal would be to generalize to finite $\mu$ previous results at $\mu=\infty$ \cite{P2014.1,P2016.2} for the spectrum of the KPZ fixed point.

High precision numerics for the Bethe roots $y_{j}$ of finite systems supplemented with extrapolation methods allow to put forward conjectures for these higher eigenstates as well. We focus again on the half-filled case $\rho=1/2$ for simplicity. The following conjectures are based on the study of a somewhat limited number of higher excited states.

We conjecture that the finite size polynomials $\hat{Q}(y)$, $\hat{P}(y)$ and $\hat{T}(y)$ scale in the same way as for the gap when $y=-1+x/\sqrt{L}$ in the large $L$ limit. One can then define corresponding functions $Q_{\mu}(x)$, $P_{\mu}(x)$ and $T_{\mu}(x)$ as in (\ref{Q[Qh]}), (\ref{P[Ph]}), (\ref{T[Th]}), which still verify the functional equations (\ref{TQ eq}), (\ref{TP eq}), (\ref{QP eq}), (\ref{TQP eq}). We also conjecture that (\ref{Q(-infty)}) is modified as $Q_{\mu}(x-\mu/2)\simeq1+4\rmi p/\mu x-8(e(\mu)+p^{2})/\mu^{2}x^{2}$ at large $|x|$, with $p$ the total momentum corresponding to an eigenvalue $\rme^{\rmi p/L}$ for the translation operator ($p=2\pi$ for the gap).

The symmetry relations (\ref{symmetry T}) and (\ref{symmetry QP}) appear to hold for higher excited states as well, at least in the finite neighbourhood of $\mu=0$ where $e(\mu)$ is real. For some higher eigenstates, $e(\mu)$ acquires an imaginary part when $\mu$ is larger than some critical value $\mu_{\text{c}}>0$, and the symmetry relations (\ref{symmetry T}), (\ref{symmetry QP}) are no longer valid beyond $\mu_{\text{c}}$.

The small $\mu$ limit (\ref{T mu 0}) of $T_{\mu}$ also seems to hold for higher excited states. This is not the case for the small $\mu$ limit (\ref{Q mu 0}) of $Q_{\mu}$: it is replaced by $Q_{\mu}(x)\sim\mu^{-m}$, with $m$ a positive integer and a coefficient equal to an unknown function. This function can in principle be obtained in a similar way as in section \ref{section direct}. A better, fully functional derivation would presumably require finding good constraints (analyticity, growth at infinity, \ldots) to supplement the functional equations in order to single out the appropriate solutions.
\end{subsection}
\end{section}

\begin{section}{Conclusions}
We studied in this paper the spectral gap $E_{1}$ of the weakly asymmetric exclusion process with hopping rates $1$ and $1-\mu/\sqrt{L}$ in the thermodynamic limit of large system size $L$ and large number of particles $N=L/2$. The spectral gap vanishes as $E_{1}\simeq e_{1}(\mu)/L^{2}$, corresponding to a rescaling of time by a factor $L^{2}$ for the time evolution of the universal model labelled by $\mu$ on the crossover between the Edwards-Wilkinson and KPZ fixed points.

Using Bethe ansatz, we obtained functional equations in the thermodynamic limit, from which a systematic perturbative expansion of $e_{1}(\mu)$ near $\mu=0$ can be computed. Relating this small $\mu$ expansion to the large $\mu$ expansion of \cite{K1995.1} is still an open question. In particular, it would be nice to describe in a non-perturbative way the solution $Q_{\mu}$ of the functional equation in the crossover between $\mu\to0$ (\ref{Q mu 0}), for which $Q_{\mu}(x)$ essentially reduces to the simple function $1+\erf(x/\sqrt{8})$, and $\mu\to\infty$ (\ref{Q mu infinity}), where the interpretation of the gap in terms of a particle-hole excitation becomes manifest. This would be especially useful for higher excited states, for which the particle-hole picture is known, but the analogue of the error function in the small $\mu$ limit is not.

At the moment, there are several gaps in our derivation of the functional equations in the thermodynamic limit, which are overcome only by high precision numerics. A careful study of the analytic properties of the finite size functional equations might provide a cleaner derivation.
\end{section}

\appendix
\begin{section}{Asymptotics of a sum}
\label{appendix asymptotics second sum f}
In this appendix, we derive the $1\ll M\ll L$ asymptotics (\ref{asymptotics sum[logj,yj]}) of the sum
\begin{equation}
S=\frac{1}{L}\sum_{j=M+1}^{N-M-1}\log_{j}\Big(\frac{1}{y}\,\frac{y-qy_{j}}{y_{j}-qy}\Big)\;.
\end{equation}
We expect from numerics that $S$ is of order $L^{-1/2}$. Since $y_{j}=f^{-1}(2\rmi\pi n_{j}/L)$ and the first correction $f_{1}$ to $f_{0}$ is of order $1/L$, replacing $f$ by $f_{0}$ from (\ref{f0}) should give the correct expansion up to order $L^{-1}$, which is what we want here. Thus $S\simeq S_{0}$ with
\begin{eqnarray}
&& S_{0}=\frac{1}{L}\sum_{j=M+1}^{N-M-1}\Big[-\rmi\arg(f_{0}^{-1}(2\rmi\pi n_{j}/L))\\
&&\hspace{29mm} +\log\Big(-\frac{\rme^{\rmi\arg(f_{0}^{-1}(2\rmi\pi n_{j}/L))}}{qy}\,\frac{y-qf_{0}^{-1}(2\rmi\pi n_{j}/L)}{y-q^{-1}f_{0}^{-1}(2\rmi\pi n_{j}/L)}\Big)\Big]\;.\nonumber
\end{eqnarray}

\begin{subsection}{Large \texorpdfstring{$L$}{L} expansion}
We first perform the large $L$ expansion with fixed $M$. We introduce
\begin{equation}
\fl\hspace{10mm} h_{L}(u)=-\rmi\arg(f_{0}^{-1}(2\rmi\pi u))+\log\Big(-\frac{\rme^{\rmi\arg(f_{0}^{-1}(2\rmi\pi u))}}{qy}\,\frac{y-qf_{0}^{-1}(2\rmi\pi u)}{y-q^{-1}f_{0}^{-1}(2\rmi\pi u)}\Big)\;,
\end{equation}
which depends on $L$ through $q=1-\mu/\sqrt{L}$ and $y=-1+w/\sqrt{L}$. It is possible to convince oneself that for all $u$ such that $-\rho/2\leq u\leq\rho/2$, one has $h_{L}(u)=\tilde{h}_{L}(u)+\mathcal{O}(L^{-3/2})$ with
\begin{eqnarray}
&&\hspace{-11mm}
  \tilde{h}_{L}(u)
  =g_{0}(L(u+\rho/2))
  +\frac{g_{1}(L(u+\rho/2))}{\sqrt{L}}
  +\frac{g_{2}(L(u+\rho/2))}{L}\nonumber\\
&&
  +g_{0}(L(u-\rho/2))
  +\frac{g_{1}(L(u-\rho/2))}{\sqrt{L}}
  +\frac{g_{2}(L(u-\rho/2))}{L}\\
&&
  -\frac{w}{\sqrt{L}}
  +\frac{\mu}{\sqrt{L}}
  \Big(
    \frac{\rme^{2\rmi\pi u}}{\sqrt{1+\rme^{4\rmi\pi u}}}
    -\frac{(\rmi\pi(u+\rho/2))^{-1/2}}{2}
    -\frac{(\rmi\pi(u-\rho/2))^{-1/2}}{2}\nonumber\\
&&\hspace{60mm}
    +\frac{\sqrt{\rmi\pi(u+\rho/2)}}{2}
    +\frac{\sqrt{\rmi\pi(u-\rho/2)}}{2}
  \Big)\nonumber\\
&&
  -\frac{w^{2}}{2L}+\frac{\mu^{2}}{L}
  \Big(
    \frac{\rme^{2\rmi\pi u}}{2\sqrt{1+\rme^{4\rmi\pi u}}}
    -\frac{(\rmi\pi(u+\rho/2))^{-1/2}}{4}
    -\frac{(\rmi\pi(u-\rho/2))^{-1/2}}{4}
  \Big)\nonumber\\
&&\hspace{11mm}
  +\frac{w\mu}{2L}
  \Big(
    \frac{\rme^{4\rmi\pi u}}{1+\rme^{4\rmi\pi u}}
    -\frac{(u+\rho/2)^{-1}}{4\rmi\pi}
    -\frac{(u-\rho/2)^{-1}}{4\rmi\pi}
  \Big)\nonumber
\end{eqnarray}
with
\begin{eqnarray}
&&
  g_{0}(\delta)=\log\frac{w-\mu-4\sqrt{\rmi\pi\delta}}{w+\mu-4\sqrt{\rmi\pi\delta}}\\
&&
  g_{1}(\delta)
  =w
  +\frac{w^{2}-\mu^{2}}{2(w-\mu-4\sqrt{\rmi\pi\delta})}
  -\frac{w^{2}+\mu^{2}}{2(w+\mu-4\sqrt{\rmi\pi\delta})}\\
&&
  g_{2}(\delta)
  =\frac{w^{2}}{2}
  -\frac{w\mu}{4}
  +\frac{\mu\sqrt{\rmi\pi\delta}}{2}\\
&&\hspace{9mm}
  +\frac{(w-\mu)(5w^{2}+6w\mu+5\mu^{2})}{16(w-\mu-4\sqrt{\rmi\pi\delta})}
  -\frac{(w+\mu)(5w^{2}-6w\mu+5\mu^{2})}{16(w+\mu-4\sqrt{\rmi\pi\delta})}\nonumber\\
&&\hspace{9mm}
  -\frac{(w^{2}-\mu^{2})^{2}}{8(w-\mu-4\sqrt{\rmi\pi\delta})^{2}}
  +\frac{(w^{2}+\mu^{2})^{2}}{8(w+\mu-4\sqrt{\rmi\pi\delta})^{2}}
  \;,\nonumber
\end{eqnarray}
and that it is sufficient to replace $h_{L}$ by $\tilde{h}_{L}$ in the summand for the expansion of $S_{0}$ up to order $L^{-1}$. Then, using the Euler-Maclaurin formulas from \cite{P2015.2} in order to treat the mix of logarithmic and square root singularities, it is possible to perform the asymptotics of each term. Gathering the contributions of all terms, we finally obtain the large $L$, fixed $M$ asymptotics of $S_{0}$:
\begin{eqnarray}
\label{S0 large L}
&&\fl\hspace{5mm}
  S_{0}\simeq
  \frac{w+\mu}{2\sqrt{L}}
  +\frac{1}{L}
  \Bigg[
    \frac{w^{2}+\mu^{2}-w\mu/2}{4}
    +\frac{1}{2}\log\frac{\rme^{\frac{(w-\mu)^{2}}{16}}-\rme^{-\frac{(w-\mu)^{2}}{16}}}{\rme^{\frac{(w+\mu)^{2}}{16}}
    -\rme^{-\frac{(w+\mu)^{2}}{16}}}\\
&&\hspace{10mm}
    +\frac{\sgn(\Im w)}{\sqrt{4\rmi\pi}}
    \Big(
      \int_{0}^{\frac{(w+\mu)^{2}}{8}}\!\!\!\!\!\rmd v\,\frac{\Li_{1/2}(-\rme^{v})}{\sqrt{\frac{(w+\mu)^{2}}{8\rmi}+\rmi v}}
      -\int_{0}^{\frac{(w-\mu)^{2}}{8}}\!\!\!\!\!\rmd v\,\frac{\Li_{1/2}(-\rme^{v})}{\sqrt{\frac{(w-\mu)^{2}}{8\rmi}+\rmi v}}
    \Big)\nonumber\\
&&\hspace{35mm}
    -\sum_{j=1}^{M+1}\log\frac{\sqrt{j-1}-\frac{w-\mu}{4\sqrt{\rmi\pi}}}{\sqrt{j-1}-\frac{w+\mu}{4\sqrt{\rmi\pi}}}
    -\sum_{j=1}^{M}\log\frac{\sqrt{j-1}-\frac{w-\mu}{4\sqrt{-\rmi\pi}}}{\sqrt{j-1}-\frac{w+\mu}{4\sqrt{-\rmi\pi}}}
  \Bigg]\;.\nonumber
\end{eqnarray}
This intermediate result was checked with high precision for small values of $M$ using Richardson extrapolation.
\end{subsection}

\begin{subsection}{Large \texorpdfstring{$M$}{M} expansion}
The second step is to extract the large $M$ limit of (\ref{S0 large L}). Euler-Maclaurin formulas from \cite{P2015.2} can be used directly. They contribute in particular terms cancelling the integrals in (\ref{S0 large L}). In the end, we obtain the announced result (\ref{asymptotics sum[logj,yj]}). The cancellation of most terms from (\ref{S0 large L}) is presumably the sign that a simpler derivation should exist.
\end{subsection}
\end{section}

\begin{section}{Sum formulas for the zeroes of \texorpdfstring{$1+\erf$}{1+erf}}
\label{appendix sum w0}
In this appendix, we prove the two sum formulas (\ref{sum 1/w2}), (\ref{sum 1/w}) for the zeroes $w_{j}(0)$, $j\in\mathbb{Z}^{*}$ of the function $\mathcal{Q}(x)=1+\erf(\frac{x}{2\sqrt{2}})$, which are represented in figure \ref{fig w0}. In both cases, we use
\begin{equation}
\label{sum w[contour integral]}
\sum_{j\in\mathbb{Z}^{*}}f(w_{j}(0))=\oint_{\mathcal{C}}\frac{\rmd z}{2\rmi\pi}\,f(z)\,\frac{\mathcal{Q}'(z)}{\mathcal{Q}(z)}\;,
\end{equation}
where the counter-clockwise contour $\mathcal{C}$ encloses the $w_{j}(0)$ but none of the singularities of $f$.

\begin{subsection}{Derivation of (\ref{sum 1/w2})}
\label{appendix sum 1/w0*w0}
In this section, we prove (\ref{sum 1/w2}). Using (\ref{sum w[contour integral]}) with $f(z)=1/z^{2}$, we observe that the portion of the contour to the right of the zeroes does not contribute since $\mathcal{Q}'(x)/\mathcal{Q}(x)$ is exponentially small in this region. Since $w_{j}(0)\sim\sqrt{\rmi j}$ when $j\to\pm\infty$, one can write
\begin{equation}
\sum_{j\in\mathbb{Z}^{*}}\frac{1}{w_{j}(0)^{2}}=
\lim_{\Lambda\to\infty}
\Big[
  \int_{\epsilon+\rmi\Lambda}^{\epsilon-\rmi\Lambda}
  +\int_{\Lambda+\rmi\Lambda}^{\epsilon+\rmi\Lambda}
  +\int_{\epsilon+\rmi\Lambda}^{\Lambda-\rmi\Lambda}
\Big]
\frac{\rmd z}{2\rmi\pi z^{2}}\,\frac{\mathcal{Q}'(z)}{\mathcal{Q}(z)}
\end{equation}
with $0<\epsilon<\Re\,w_{\pm1}$. The last two integrals can be computed explicitly using the large $z$ expansion $\mathcal{Q}'(z)/\mathcal{Q}(z)\simeq-z/4$ outside $\{z\in\mathbb{C},-\frac{\pi}{4}<\arg z<\frac{\pi}{4}\}$. This leads to
\begin{equation}
\sum_{j\in\mathbb{Z}^{*}}\frac{1}{w_{j}(0)^{2}}=-\frac{1}{16}+\int_{\epsilon+\rmi\infty}^{\epsilon-\rmi\infty}\frac{\rmd z}{2\rmi\pi z^{2}}\,\frac{\mathcal{Q}'(z)}{\mathcal{Q}(z)}\;.
\end{equation}
The contour in the last integral can then be shifted to the left after computing the residue at $0$, leading to
\begin{equation}
\sum_{j\in\mathbb{Z}^{*}}\frac{1}{w_{j}(0)^{2}}=\frac{1}{2\pi}-\frac{1}{16}+\int_{-\epsilon+\rmi\infty}^{-\epsilon-\rmi\infty}\frac{\rmd z}{2\rmi\pi z^{2}}\,\frac{\mathcal{Q}'(z)}{\mathcal{Q}(z)}\;.
\end{equation}
Sending $\epsilon$ to infinity, one can use $\mathcal{Q}'(z)/\mathcal{Q}(z)\simeq-z/4$ again to compute the remaining integral, which gives finally (\ref{sum 1/w2}).
\end{subsection}

\begin{subsection}{Derivation of (\ref{sum 1/w})}
In this section, we prove (\ref{sum 1/w}). Using the restriction of (\ref{sum w[contour integral]}) with $f(z)=1/z$ to the first $2M$ zeroes, the contour to the right of the zeroes does not contribute again, and one has
\begin{eqnarray}
&&\hspace{-25mm}
\lim_{M\to\infty}\sum_{j\in[\![-M,M]\!]^{*}}\Big(\frac{1}{w_{j}(0)}-\frac{1}{4\sqrt{\rmi\pi j}}\Big)
=\lim_{M\to\infty}
\Big(
  -\sum_{j\in[\![-M,M]\!]^{*}}\frac{1}{4\sqrt{\rmi\pi j}}\\
&&\hspace{25mm}
+\Big[
  \int_{\epsilon+\rmi\,\Im\,w_{M}}^{\epsilon+\rmi\,\Im\,w_{-M}}
  +\int_{w_{M}}^{\epsilon+\rmi\,\Im\,w_{M}}
  +\int_{\epsilon+\rmi\,\Im\,w_{-M}}^{w_{-M}}
\Big]
\frac{\rmd z}{2\rmi\pi z}\,\frac{\mathcal{Q}'(z)}{\mathcal{Q}(z)}
\Big)\;.\nonumber
\end{eqnarray}
The last two integrals can be computed explicitly for large $M$ using $\mathcal{Q}'(z)/\mathcal{Q}(z)\simeq-z/4$. Their contributions cancel. Then, the contour of the first integral can be shifted to the left after computing the residue at $0$. We find
\begin{eqnarray}
&&\hspace{-25mm}
\lim_{M\to\infty}\sum_{j\in[\![-M,M]\!]^{*}}\Big(\frac{1}{w_{j}(0)}-\frac{1}{4\sqrt{\rmi\pi j}}\Big)
=\lim_{M\to\infty}
\Big(
  -\frac{1}{\sqrt{2\pi}}
  -\sum_{j\in[\![-M,M]\!]^{*}}\frac{1}{4\sqrt{\rmi\pi j}}\\
&&\hspace{75mm}
  +\int_{-\epsilon+\rmi\,\Im\,w_{M}}^{-\epsilon+\rmi\,\Im\,w_{-M}}\frac{\rmd z}{2\rmi\pi z}\,\frac{\mathcal{Q}'(z)}{\mathcal{Q}(z)}
\Big)\;.\nonumber
\end{eqnarray}
Sending $\epsilon$ to $-\infty$, one can use again $\mathcal{Q}'(z)/\mathcal{Q}(z)\simeq-z/4$ to compute the last integral. The final result (\ref{sum 1/w}) follows from $w_{\pm M}\simeq4\sqrt{\pm\rmi\pi M}$ and $\sum_{j=1}^{M}j^{-1/2}\simeq2\sqrt{M}+\zeta(1/2)$ at large $M$.
\end{subsection}
\end{section}

\vspace{10mm}

\end{document}